\documentclass[conference]{IEEEtran}

\let\labelindent\relax
\usepackage{enumitem}
\usepackage{etex}
\usepackage{amsmath,amssymb,amsfonts}
\usepackage{algorithmic}
\usepackage{graphicx}
\expandafter\def\csname ver@subfig.sty\endcsname{}
\usepackage{textcomp}
\usepackage[table]{xcolor}
\usepackage{epsfig,endnotes}
\usepackage{opera}
\usepackage{color}
\usepackage{hyperref}
\usepackage{url}
\usepackage{pgfplots, pgfplotstable}
\usepgfplotslibrary{groupplots}
\usepackage[htt]{hyphenat}
\usepackage{pifont}
\usepackage[linesnumbered,lined,vlined,ruled,commentsnumbered,noend]{algorithm2e}
\SetAlCapNameFnt{\scriptsize}
\SetAlCapFnt{\scriptsize}
\usepackage{tabulary}
\usepackage{booktabs}
\usepackage{bbm}
\usepackage{tikz}
\usetikzlibrary{matrix,positioning}
\usepackage{multirow}
\usepackage{multicol}
\usepackage{placeins}
\usepackage{comment}
\usepackage{mdframed}
\usepackage{soul}
\usepackage[font={footnotesize}, skip=0pt]{caption}
\def\BibTeX{{\rm B\kern-.05em{\sc i\kern-.025em b}\kern-.08em
    T\kern-.1667em\lower.7ex\hbox{E}\kern-.125emX}}
\usepackage{subfigure}
\usepackage[noadjust]{cite}

\begin{document}

\newtheorem{mydefinition}{Definition}
\newcommand{\esp}{esp.\@\xspace}
\newcommand{\projectname}{\texttt{CamQuery}\xspace}
\newcommand{\camflow}{\texttt{CamFlow}\xspace}
\newcommand{\unicorn}{\textsc{Unicorn}\xspace}
\newcommand{\cL}{{\cal L}}
\renewcommand{\figureautorefname}{Fig.\footnotesize}
\renewcommand{\tableautorefname}{Table}
\renewcommand{\algorithmautorefname}{Alg.}
\renewcommand{\sectionautorefname}{\S}
\renewcommand{\subsectionautorefname}{\S}
\renewcommand{\subsubsectionautorefname}{\S}
\providecommand*{\lstlistingautorefname}{Listing}

\newcommand*\circled[1]{\tikz[baseline=(char.base)]{
            \node[shape=circle,draw,inner sep=0pt] (char) {{\small#1}};}}

\title{\unicorn: Runtime Provenance-Based Detector for Advanced Persistent Threats}

\newmdtheoremenv[
hidealllines=true,
leftline=true,
innertopmargin=0pt,
innerbottommargin=0pt,
linewidth=2pt,
linecolor=gray!40,
innerrightmargin=0pt,
]{definitionii}{Definition}

\IEEEoverridecommandlockouts
\makeatletter\def\@IEEEpubidpullup{6.5\baselineskip}\makeatother
\IEEEpubid{\parbox{\columnwidth}{
		Network and Distributed Systems Security (NDSS) Symposium 2020\\
		23-26 February 2020, San Diego, CA, USA\\
		ISBN 1-891562-61-4\\
		https://dx.doi.org/10.14722/ndss.2020.24046\\
		www.ndss-symposium.org
	}
\hspace{\columnsep}\makebox[\columnwidth]{}}

\author{\IEEEauthorblockN{Xueyuan Han\IEEEauthorrefmark{1},
Thomas Pasquier\IEEEauthorrefmark{2},
Adam Bates\IEEEauthorrefmark{3}, 
James Mickens\IEEEauthorrefmark{1} and
Margo Seltzer\IEEEauthorrefmark{4}}
\IEEEauthorblockA{\IEEEauthorrefmark{1}Harvard University\\
\{hanx,mickens\}@g.harvard.edu}
\IEEEauthorblockA{\IEEEauthorrefmark{2}University of Bristol\\
thomas.pasquier@bristol.ac.uk}
\IEEEauthorblockA{\IEEEauthorrefmark{3}University of Illinois at Urbana-Champaign\\
batesa@illinois.edu}
\IEEEauthorblockA{\IEEEauthorrefmark{4}University of British Columbia\\
mseltzer@cs.ubc.ca}}

\maketitle

\begin{abstract}
Advanced Persistent Threats (APTs) are difficult to detect 
due to their ``low-and-slow'' attack patterns and frequent use of zero-day exploits.
We present \unicorn,
an anomaly-based APT detector that effectively leverages data provenance analysis.
From modeling to detection,
\unicorn tailors its design specifically for the unique characteristics of APTs.
Through extensive yet time-efficient graph analysis,
\unicorn explores provenance graphs that provide rich contextual and historical information
to identify stealthy anomalous activities without pre-defined attack signatures.
Using a graph sketching technique,
it summarizes long-running system execution with space efficiency
to combat slow-acting attacks that take place over a long time span.
\unicorn further improves its detection capability 
using a novel modeling approach to understand long-term behavior as the system evolves.
Our evaluation shows that \unicorn
outperforms an existing state-of-the-art APT detection system
and detects real-life APT scenarios with high accuracy.
 \end{abstract}

\maketitle

\section{Introduction}
\label{sec:introduction}
Advanced Persistent Threats (APT) are becoming increasingly common~\cite{alshamrani2019survey}.
The long timescale over which such attacks take place makes them fundamentally
different from more conventional attacks.
In an APT, the adversary's goal is to gain control of a specific (network of) system(s) 
while remaining undetected for an extended period of time~\cite{symantec2011advanced}.
The adversary often relies on zero-day exploits~\cite{nikos2014big, symantec2011advanced} to gain a foothold in the victim system.

Traditional detection systems are not well-suited to APTs.
Detectors dependent on malware signatures are blind to attacks that
exploit new vulnerabilities~\cite{bilge2012before}.
Anomaly-based systems typically analyze series of 
system calls~\cite{somayaji2000automated} and log-adjacent system events~\cite{xu2016sharper},
but most of them~\cite{sekar2001fast,feng2003anomaly,mutz2007exploiting,maggi2010detecting}
have difficulty modeling long-term behavior patterns.
Further, they are susceptible to evasion techniques,
because they typically inspect only short sequences of system calls and events. 
As a result, they have exhibited little success in detecting APTs.
Systems that attempt to capture long-term program behavior~\cite{shu2017long}
limit their analysis to event co-occurrence to avoid high computational and memory overheads.

More recent work~\cite{milajerdi2019holmes,manzoor2016fast,berrada2019aggregating,barre2019mining} suggests that data provenance might be a better data source
for APT detection.
Data provenance represents system execution as a directed acyclic graph (DAG)
that describes information flow between system subjects (\eg processes) and objects (\eg files and sockets).
It connects causally-related events in the graph, even when those events
are separated by a long time period.
Thus, even though systems under APT attack usually behave similarly to
unattacked systems, the richer contextual information in provenance
allows for better separation of benign and malicious events~\cite{hassannodoze}.

However,
leveraging data provenance to perform runtime APT analysis is difficult.
Provenance graph analysis is computationally demanding, because
the graph size grows continuously as APTs slowly penetrate a system.
The necessary contextual analysis, which requires large graph components,
makes the task even more challenging.
One approach to provenance-based APT detection~\cite{milajerdi2019holmes} 
uses simple edge-matching rules based on prior attack knowledge,
but this makes it difficult to detect new classes of APTs~\cite{barre2019mining}.
Provenance-based anomaly detection systems
rely on graph neighborhood exploration
to understand normal behavior through either static~\cite{han2017frappuccino} or dynamic~\cite{manzoor2016fast} models.
However, practical computational constraints limit the scope of the
contextual analysis that is feasible.
Thus, current systems suffer from some combination of the following three
problems:
1) static models cannot capture long term system behavior,
2) the low-and-slow APT approach can gradually poison dynamic models, and
3) approaches that require in-memory computation~\cite{milajerdi2019holmes,manzoor2016fast} scale poorly in the presence of long-running attacks~\cite{milajerdi2019holmes}.

We introduce \unicorn,
a provenance-based anomaly detector capable of detecting APTs.
\unicorn uses graph sketching to build an incrementally updatable, fixed size, longitudinal graph data structure
\footnote{An algorithm that builds a graph data structure is called a \emph{graph kernel}, which
is an overloaded term that also refers to functions that compare the similarity between two graphs.
We adopt the second definition here.}
that enables efficient computation of graph statistics~\cite{fairbanks2013statistical}.
This longitudinal nature of the structure
permits extensive graph exploration, 
which allows \unicorn to track stealthy intrusions.
The fixed size and incrementally updatable graph data structure obviates the 
need for an in-memory representation of the provenance graph, so
\unicorn is scalable with low computational and storage overheads.
It can succinctly track
a machine's entire provenance history from boot to shutdown.
\unicorn directly models a system's evolving behavior during training,
but it does not update models afterward,
preventing an attacker from poisoning the model.

\noindent We make the following contributions:
\begin{itemize}[leftmargin=*]
	\setlength\itemsep{0em}
	\item We present a provenance-based anomaly detection system tailored to APT attacks.
	\item We introduce a novel sketch-based, time-weighted provenance encoding that is compact and able to usefully summarize provenance graphs over long time periods.
	\item We evaluate \unicorn against both simulated and real-life APT attack scenarios. 
	We show that \unicorn can detect APT campaigns with high accuracy.
	Compared to previous work,
	\unicorn significantly improves precision and accuracy by 24\% and 30\%, respectively.
	\item We provide an open source implementation.
\end{itemize}
\begin{comment}
	\item It follows an anomaly-based design to detect zero-day exploits. 
	\item \unicorn continuously monitor streaming provenance graph. It does not sacrifice detection capability for space and memory efficiency. Instead, it keeps in mind those constraints while maintaining its capability to explore provenance graphs and perform graph analysis.
	\begin{itemize}
			\item It efficiently takes into consideration large neighborhoods for contextualized threat analysis.
			\item It is able to take multi-label attributes in the graph.
			\item It does not keep the entire graph in-memory for computation.
	\end{itemize}
	\item	Given the continuous nature of system execution and monitoring, \unicorn directly models the dynamic behavior of the system without using static snapshots. Once the dynamic model is built, it does not attempts to modify the model as the system progresses in realtime deployment. That is, \unicorn maximizes the expressiveness/flexibility of the model when it is guaranteed to learn only benign behavior, and avoids later model changes that risk ``low-and-slow'' attacks stealthily ``taking over" the model.
\end{comment}

\section{Background}
\label{sec:background}
Traditional anomaly-based intrusion detection systems (IDS)
analyze system call traces from audit systems~\cite{forrest1996sense, ye2000markov, maggi2010detecting, xu2016sharper}. 
However, for APT detection,
whole-system provenance~\cite{pasquier2017practical} is a superior data source.

\subsection{Challenges of Syscall Traces}
\label{sec:background:syscall}

The system call abstraction provides a simple interface by which user-level applications request services of the operating system.
As the mechanism through which system services are invoked, the system call interface is also generally the entry point for attackers trying to subvert a system~\cite{jaeger2004consistency}.
Therefore, system call traces have long been regarded as the \emph{de facto} information source for intrusion detection~\cite{forrest1996sense}.

However,
existing systems capture unstructured collections of syscall audit logs,
requiring analysis to make sense of such information.
Given the low-and-slow nature of APTs,
analyzing individual syscall logs to detect point-wise outliers is often fruitless~\cite{chandola2009anomaly},
as is inspection of short system call sequences.
Such analysis does not reflect the historical context of each syscall event,
resulting in high false positive rates and mimicry attacks that evade detection~\cite{wagner2002mimicry, parampalli2008practical}.
In contrast, data provenance encodes historical context into causality graphs~\cite{carata2014primer}.

Data provenance can be used to model a variety of event sequences in computing, 
including syscall audit logs. 
Indeed, there exist frameworks that reconstruct provenance graph structures 
from audit data streams to allow for better reasoning about system execution~\cite{gehani2012spade}.
However, such post-hoc approaches make it difficult to ensure graph correctness~\cite{pohly2012hi};
it is difficult to prove completeness, trustworthiness, or reliability of the syscall traces from which the graph is built,
because many syscall interposition techniques suffer from concurrency issues~\cite{garfinkel2003traps, watson2007exploiting}.
It is easy to bypass library-wrapper-based syscall capture mechanisms~\cite{jain2000user},
while user-space mechanisms (\eg \texttt{ptrace}) incur unacceptable runtime performance overheads~\cite{garfinkel2003traps}
and are susceptible to race conditions~\cite{jain2000user}.
The same race condition issues also plague in-kernel mechanisms (\eg \texttt{Systrace}~\cite{provos2006systrace}, Janus~\cite{goldberg1996secure}),
resulting in time-of-check-to-time-of-use (TOCTTOU), time-of-audit-to-time-of-use (TOATTOU), 
and time-of-replacement-to-time-of-use (TORTTOU) bugs~\cite{watson2007exploiting}.
For example, 
many IDS~\cite{maggi2010detecting, xu2016sharper} analyze syscalls and their
arguments to defend against mimicry attacks,
but TOATTOU bugs cause the captured syscall arguments to be different
from the true values accessed in the kernel~\cite{garfinkel2003traps}.
Perhaps more importantly, instead of a single graph of system execution, 
syscall-based provenance frameworks produce many disconnected graphs,
because they cannot trace the interrelationships of kernel threads that do not make use of the syscall interface.
Consequently, such frameworks can rarely detect the stealthy malicious events
found in APTs.

\subsection{Whole-System Provenance}
\label{sec:background:whole}

Whole-system provenance collection runs at the operating system level, capturing all system activities and the interactions between them~\cite{pohly2012hi}.
OS-level provenance systems such as Hi-Fi~\cite{pohly2012hi}, 
LPM~\cite{bates2015trustworthy}, 
and CamFlow~\cite{pasquier2017practical}
provide strong security and completeness guarantees with respect to information flow capture.
This completeness is particularly desirable in APT scenarios
as it captures long-distance causal relationships enabling
contextualized analysis,
even if a malicious agent manipulates security-sensitive kernel
objects to hide its presence. 

We use CamFlow~\cite{pasquier2017practical} as the reference implementation throughout the paper, although there
exist other whole-system provenance implementations;
in \autoref{sec:evaluation}, we show that \unicorn works seamlessly
with other capture mechanisms as well.
CamFlow adopts the Linux Security Modules (LSM) framework~\cite{morris2002linux} to ensure high-quality, 
reliable recording of information flows among data objects~\cite{georget2017verifying,pasquier2018ccs}.
LSM eliminates race conditions (\eg TOCTTOU attacks) by placing mediation points inside the kernel
instead of at the system call interface~\cite{jaeger2004consistency}.

\subsection{Summary and Problem Statement}
\label{sec:background:statement}
\colorlet{soulred}{red!20}
\sethlcolor{soulred}

Prior research~\cite{milajerdi2019holmes,manzoor2016fast,barre2019mining} explored the use of data provenance for APT detection.
However, these approaches all suffer from some combinations of the
following limitations:
\begin{itemize}[leftmargin=*]
	\setlength\itemsep{0em}
	\item[\circled{L1}:] Pre-defined edge-matching rules are overly sensitive and make it difficult to detect zero-day exploits common in APTs~\cite{milajerdi2019holmes}.
	\item[\circled{L2}:] Constrained provenance graph exploration provides only limited understanding of information context critical to detect low-profile anomalies. 
	For example, graph exploration is limited to small graph neighborhoods, single node/edge attributes, and truncated subgraphs~\cite{han2017frappuccino,manzoor2016fast,barre2019mining}.
	\item[\circled{L3}:] System behavior models fail to cater to the unique characteristics of APT attacks. Static models cannot capture dynamic behavior of long-running systems~\cite{han2017frappuccino}, while
	dynamic modeling during runtime risks poisoning from the attackers~\cite{manzoor2016fast}.
	\item[\circled{L4}:] Provenance graphs are stored and analyzed only in memory, sacrificing long-term scalability~\cite{manzoor2016fast, milajerdi2019holmes}. 
\end{itemize}

\unicorn addresses those issues.
We formalize the system-wide intrusion detection problem in APT campaigns as a real-time graph-based anomaly detection problem
on large, attributed, streaming whole-system provenance graphs.
At any point in time,
the entirety of a provenance graph,
captured from system boot to its current state is compared against a behavior model consisting of known good provenance graphs.
The system is considered under attack if its provenance graph deviates significantly from the model.
For APT detection, an ideal provenance-based IDS must:
\begin{itemize}[leftmargin=*]
	\setlength\itemsep{0em}
	\item Continuously analyze provenance graphs with space and time efficiency while taking full advantage of rich information content provided by whole-system provenance graphs;
	\item Take into consideration the entire duration of system execution without making assumptions of attack behavior;
	\item Learn only normal system behavior changes but not those directed by the attackers.
\end{itemize}
  
\section{Threat Model}
\label{sec:threat}
We assume APT scenarios for host intrusion detection:
  an attacker illegitimately gains access to a system
  and plans to remain there for an extended period of time without being detected.
The attacker may conduct the attack in several phases
 and use a variety of techniques during each phase~\cite{yadav2015technical}.
The goal of \unicorn is to detect such attacks at any stage
  by interpreting the provenance generated by the host.
We assume that, prior to the attack,
  \unicorn thoroughly observes the host system during normal operation
  and that no attacks arise during this initial modeling period.

The integrity of the data collection framework is central to \unicorn{'s}
correctness.
We assume that Linux Security Modules~\cite{morris2002linux}, 
   which is the Linux reference monitor implementation, 
   correctly provides reference monitor guarantees~\cite{a1972} for CamFlow~\cite{pasquier2017practical}.
Specifically, we assume that LSM integrity is provided via an attested boot sequence~\cite{bates2015trustworthy}.
 We make similar integrity assumptions for other data collection frameworks. 
\unicorn 
  can further safeguard its data source by streaming it across the network.
  While we primarily envision \unicorn as an endpoint security monitor,
  \unicorn's ability to stream provenance data enables off-host intrusion detectors
  that are not co-located with a potentially compromised machine.

For the remainder of the paper, we assume the correctness of the kernel,
the provenance data, and the analysis engine.
We instead focus on \unicorn's analytic capabilities.
 
\section{Design}
\label{sec:design}
\begin{figure}[t]
	\centering
	\includegraphics[width=\columnwidth]{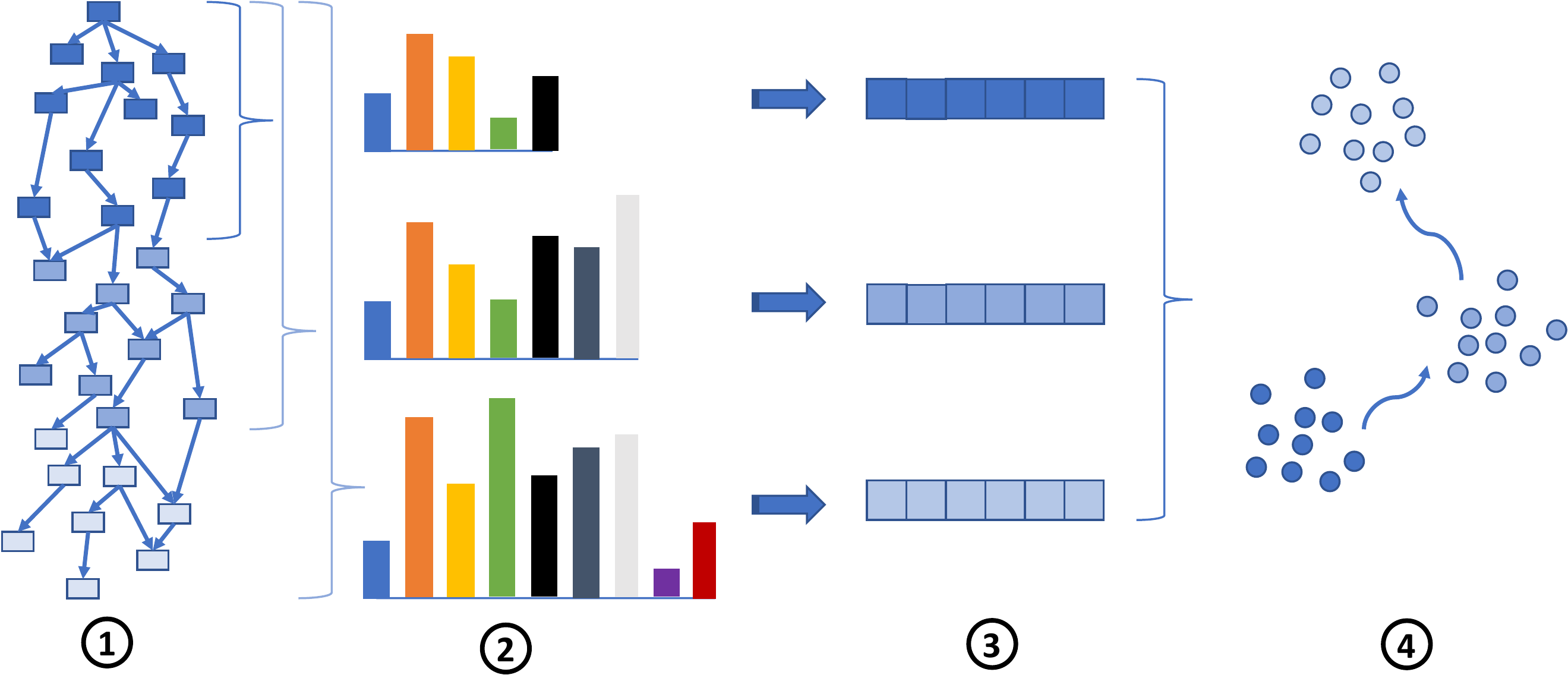}
	\caption{\unicorn \protect\circled{1} takes a streaming provenance graph, \protect\circled{2} periodically summarizes graph features
	into histograms, and then \protect\circled{3} creates fixed-size graph sketches.
	The resulting clustering-based model \protect\circled{4} captures the dynamics of system execution.
	During deployment, graph sketches are created through the same steps (\protect\circled{1}, \protect\circled{2} and \protect\circled{3})
	and then compared against the model in \protect\circled{4}.}
	\label{img:pipeline}
\end{figure}
\unicorn is a host-based intrusion detection system capable of simultaneously
detecting intrusions on a collection of networked hosts.
We begin with a brief overview of \unicorn
and then follow with a detailed discussion
of each system component in the following sections.
\autoref{img:pipeline} illustrates \unicorn's general pipeline.

\noindgras{\circled{1} Takes as input a labeled, streaming provenance graph.}
\unicorn accepts a stream of attributed edges produced by a provenance
capture system running on one or more networked hosts.
Provenance systems construct a single, whole-system provenance DAG with a partial-order guarantee, 
which allows for efficient streaming computation (\autoref{sec:design:histogram}) and fully contextualized analysis (\circled{L2}).
We present \unicorn using CamFlow~\cite{pasquier2017practical},
although it can obtain provenance from other systems,
such as LPM~\cite{bates2015trustworthy} and Spade~\cite{gehani2012spade},
the latter of which interoperates with commodity audit systems such as Linux Audit and Windows ETW.

\noindgras{\circled{2} Builds at runtime an in-memory histogram.}
\unicorn efficiently constructs a streaming graph histogram that represents the entire history of system execution,
updating the counts of histogram elements as new edges arrive in the graph data stream.
By iteratively exploring larger graph neighborhoods,
it discovers causal relationships between system entities providing execution context.
This is \unicorn's first step in building an efficient data structure
that facilitates contextualized graph analysis (\circled{L2}).
Specifically,
each element in the histogram describes a unique substructure of the graph,
taking into consideration the heterogeneous label(s) attached to the vertices and edges within the substructure,
as well as the temporal order of those edges.

To adapt to expected behavioral changes during the course of normal system execution,
 \unicorn periodically discounts the influence of histogram elements that have no causal relationships with recent events (\circled{L3}).
Slowly ``forgetting'' irrelevant past events allows us to effectively model meta-states (\autoref{sec:design:model}) throughout system uptime 
(\eg system boot, initialization, serving requests, failure modes, \etc).
However, it does not mean that \unicorn forgets informative execution history;
rather,
\unicorn uses information flow dependencies in the graph to keep up-to-date important, relevant context information.
Attackers can slowly penetrate the victim system in an APT,
hoping that a time-based IDS eventually forgets this initial attack,
but they cannot break the information flow dependencies that are essential to the success of the attack~\cite{milajerdi2019holmes}.

\noindgras{\circled{3} Periodically, computes a fixed-size graph sketch.}
In a pure streaming environment,
the number of unique histogram elements can grow arbitrarily large as
\unicorn summarizes the entire provenance graph.
This variation in size makes
it challenging to efficiently compute similarity between two histograms
and impractical to design algorithms for later modeling and detection.
\unicorn employs a similarity-preserving hashing technique~\cite{yang2017histosketch}
to transform the histogram to a \emph{graph sketch}~\cite{ahn2012graph}.
The graph sketch is incrementally maintainable, meaning that \unicorn
does not need to keep the entire provenance graph in memory;
its size is constant (\circled{L4}).
Additionally,
graph sketches preserve normalized Jaccard similarity~\cite{ji2013min} between two graph histograms.
This distance-preserving property is particularly important to the clustering algorithm in our later analysis, 
which is based on the same graph similarity metric.

\noindgras{\circled{4} Clusters sketches into a model.}
\unicorn builds a normal system execution model and identifies abnormal activities without attack knowledge (\circled{L1}).
However,
unlike traditional clustering approaches, 
\unicorn takes advantage of its streaming capability to generate models that are \emph{evolutionary}.
The model captures behavioral changes within a single execution by clustering system activities at various stages of its execution,
but \unicorn does not modify models dynamically during runtime when the attacker may be
subverting the system (\circled{L3}).
It is therefore more suitable for long-running systems under potential APT attacks.

\subsection{Provenance Graph}
\label{sec:design:graph}
Provenance graphs are increasingly popular for attack analysis~\cite{pei2016hercule, gao2018saql, liu2018towards, barre2019mining} and are
attractive for APT detection.
In particular, provenance graphs capture causality relationships between events.
Causal connectivity facilitates reasoning over events that are temporally distant, 
thus useful in navigating through APT's low-and-slow attack pattern.
Audit log analysis frequently relies on temporal relationships, while
provenance analysis leverages causality relationships, producing a more
meaningful model of system behavior.

\unicorn compares two system executions based on the similarity between their corresponding provenance graphs.
\unicorn always considers the entire provenance graph
to detect long-running attacks.
There exist many graph similarity measures,
but many approaches (\eg graph isomorphism) are too restrictive (\ie require two graphs to be exactly or largely identical)~\cite{shervashidze2011weisfeiler},
because even normal executions often produce slightly different provenance graphs.
Whole-system provenance graphs can grow large quickly~\cite{pasquier2017practical}, so
NP-complete~\cite{garey1979guide, neuhaus2005self} and even polynomial algorithms~\cite{kondor2009graphlet, vishwanathan2010graph}
are too computationally expensive for streaming settings.
As we show in the following sections,
\unicorn's graph similarity algorithm does not exhibit these problems.
 
\subsection{Constructing Graph Histograms}
\label{sec:design:histogram}
Our goal is to efficiently compare provenance graphs
while tolerating minor variations in normal executions.
The two criteria we have for an algorithm are:
1) the representation should take into account
long-term causal relationships, and
2) we must be able to implement the algorithm on the real-time streaming graph
data so that we can thwart intrusions when they happen (not merely detect them).

We adapt a linear-time, fast Weisfeiler-Lehman (WL) subtree graph kernel algorithm based on one dimensional WL test of isomorphism~\cite{weisfeiler1968reduction}.
The WL test of isomorphism and its subtree kernel variation~\cite{shervashidze2011weisfeiler} are known for their discriminative power for a broad class of graphs,
beyond many state-of-the-art graph learning algorithms (\eg graph neural networks~\cite{hamilton2017inductive, xu2018powerful}).

Our use of the WL subtree graph kernel hinges on our ability to construct
a histogram of vertices that captures the graph structure surrounding each
vertex.
We bin vertices according to augmented vertex labels that describe fully the
$R$-hop neighborhood of the vertex.
We construct these augmented vertex labels by iterative label propagation;
we provide an intuitive description here and a more formal one below.
For simplicity of exposition, assume we have an entire static graph.
A single relabeling step takes as input a vertex label, the labels of all
its incoming edges, and the labels of the source vertices of all those
edges.
It then outputs a new label for the vertex representing the aggregation of
all the input labels.
We repeat this process for every vertex, and then repeat the entire procedure
$R$ times to construct labels describing an $R$-hop neighborhood.
Once we have constructed augmented vertex labels for every vertex in the
graph, we create a histogram whose buckets correspond to these labels.
The WL test of isomorphism compares two graphs based on these
augmented vertex labels;
two graphs are similar if they have similar distributions across
similar sets of labels.

\begin{algorithm}[!ht]
	\algsetup{linenosize=\tiny}
	\scriptsize
	\SetAlgoLined
	\DontPrintSemicolon
	\SetKwInOut{Input}{Input}\SetKwInOut{Output}{Output}
	\SetKwData{S}{s}
	\SetKwFunction{Sort}{Sort}\SetKwFunction{In}{In}\SetKwFunction{Source}{Source}\SetKwFunction{Concat}{Concat}
	\SetKwFunction{N}{$\mathcal{N}$}\SetKwFunction{T}{$T$}\SetKwFunction{Min}{Min}\SetKwFunction{Hash}{Hash}
	\Input{$G=(V, E, \mathcal{F}^v, \mathcal{F}^e, \mathcal{C}), R$}
	\Output{Histogram $H$}
	\For{$i \leftarrow 1$ \KwTo $R$}{\label{algo:histogram:r}
		\ForEach{$v \in V$}{
			$M \leftarrow \{ \}$\;
			\If{$i == 1$}{
				$l_0(v) \leftarrow \mathcal{F}^v(v)$\;
			}
			\ElseIf{$i == 2$}{
				$TS \leftarrow \{ \}$\;
				\ForEach{$e \in \In{v}$}{\label{algo:histogram:in}
					$w$ $\leftarrow \Source{e}$\; \label{algo:histogram:source}
					$M \leftarrow M + \{\mathcal{F}^e(e)::l_1(w)\}$\;
					\T{$w$} $\leftarrow \mathcal{C}(e)$\;
					$TS \leftarrow TS + \{\T{$w$}\}$\;
				}
				\T{$v$} $\leftarrow \Min{TS}$\;
			}
			\Else{
				$TS \leftarrow \{ \}$\;
				\ForEach{$w \in \N{v}$}{\label{algo:histogram:n}
					$M \leftarrow M + \{l_{i - 1}(w)\}$\;
					$TS \leftarrow TS + \{\T{$w$}\}$\;
				}
				\T{$w$} $\leftarrow \Min{TS}$\;
			}
			\Sort{$M$} based on timestamps \T{$w$}, $\forall w$ whose label is included in $M$\;
			$\S_v$ $\leftarrow l_{i-1}(v) + \Concat{M}$\;
		}
		\ForEach{$v \in V$} {
			$l_{i}$($v$) $\leftarrow$ \Hash{$\S_v$}\; \label{algo:histogram:label}
			\lIf{$l_{i}(v) \notin H$}{$H[l_{i}(v)] \leftarrow 1$}
			\lElse{$H[l_{i}(v)] \leftarrow H[l_{i}(v)] + 1$}
		}
	}
	\caption{Graph Histogram Generation}\label{algo:histogram}
\end{algorithm}

\autoref{algo:histogram} presents the algorithm more formally.
We define a static graph $G$ as a 5-tuple $(V, E, \mathcal{F}^v, \mathcal{F}^e, \mathcal{C})$,
where $V$ is the set of vertices, $E$ is the set of directed edges, \ie $e = (u, v) \in E$ and $(u, v) \neq (v, u)$ $\forall$ $u, v \in V$.
$\mathcal{F}^v: V \rightarrow \Sigma$ is a function that assigns labels from an alphabet $\Sigma$ to each vertex in the graph,
and similarly, $\mathcal{F}^e: E \rightarrow \Psi$ assigns labels from an alphabet $\Psi$ to each edge.
$\mathcal{C}$ is a function that records the timestamp of each edge.
In \autoref{algo:histogram:r},
$R$ is the number of neighborhood hops that each vertex explores to generate histogram elements.
\unicorn creates a histogram representation of an entire graph
by examining rooted subtrees around every vertex in the graph.
By considering such non-linear substructures,
in addition to the attributes of each vertex/edge,
\unicorn preserves structural equivalence of provenance graphs,
which has been demonstrated to outperform linear approaches such as random walk~\cite{Narayanan2017graph2vecLD}.
$M$ is a list of neighboring vertex and/or edge labels of a vertex $v$
and $l_{i}(v)$ is the label (\ie histogram element) of the vertex $v$ in the $i^{th}$ iteration (or the $(i-1)$-hop neighborhood, where the $0$-hop neighborhood is the vertex itself).
$TS$ is a list of timestamps of the neighboring vertices.
In \autoref{algo:histogram:in}, the function \texttt{In}$(v)$ returns all the incoming edges of the vertex $v$
and the function \texttt{Source}$(e)$ in \autoref{algo:histogram:source} returns the source vertex of the edge $e$.
$T$ records the timestamp of each vertex $v$.
In \autoref{algo:histogram:n}, $\mathcal{N}(v) = \{w\mid (w, v) \in E\}$ is the set of vertices to which $v$ is connected by an in-coming edge.

Our goal is to construct histograms where each element of the histogram corresponds to a unique vertex label, which
captures the vertex's $R$-step in-coming neighborhood.
Labels capture information about the edges in the neighborhood and the identities of the vertices in that neighborhood,
including complex contextual information that reveals causal relationships among objects, subjects, and activities~\cite{Narayanan2017graph2vecLD}.
We bootstrap the labels as follows:
each vertex begins with its own initial label.
We then incorporate 1-hop neighbors by adding the labels of the incoming edges and the initial
vertex labels of the sources of those edges.
After this bootstrapping process, every vertex label now represents both vertex and edge labels so that as we expand to increasingly large neighborhoods, we need only add labels for the sources of all the incoming edges (since those labels already incorporate edge labels from their incoming edges).
We sort all the labels by the timestamp of their corresponding edge, respecting the sequence of events in the system,
and then hash the label list to facilitate fast lookup and bookkeeping.
Some vertices/edges may have multiple labels,
but the same hashing trick permits multi-label computation at negligible cost.

\begin{figure}[t]
	\centering
	\includegraphics[width=0.4\columnwidth, scale=0.2]{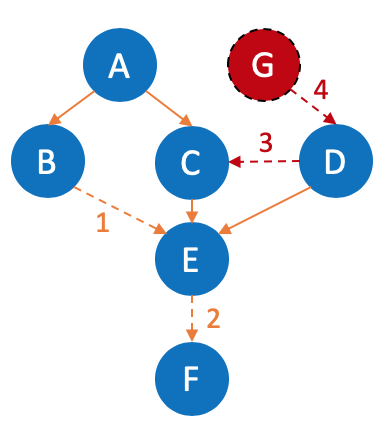}
	\caption{A simple, abstracted provenance graph where dotted edges arrive after all solid edges have been processed and dotted vertices are newly-arrived.
	If the provenance graph observes partial ordering,
	edge $1$ must arrive before edge $2$ and only the local structures of vertex $E$ and $F$ need to be computed.
	However, in a provenance graph where partial ordering is \emph{not} observed, we can encounter edge $3$ and $4$ and the newly-arrived vertex $G$ (in dotted red) after the solid edges.
	In such a case, with $R$ = 2, we need to update both vertex $C$ and $E$ (descendants of $D$) for edge $3$ and vertex $C$, $D$, and $E$ (descendants of $G$) for edge $4$.
 	}
	\label{img:example}
\end{figure}

\noindgras{Streaming Variant and Complexity.}
In a streaming environment,
we run \autoref{algo:histogram} only on newly-arriving vertices and on vertices whose in-coming neighborhood is affected by new edges.
In provenance graphs that use multiple vertices per provenance entity or activity to represent different versions or states of the corresponding object
~\cite{muniswamy2006provenance},
we need to compute/update \emph{only} the neighborhood of the destination vertex for each new edge,
because all incoming edges to a vertex arrive before any outgoing ones~\cite{pasquier2018ccs}.
\unicorn takes advantage of this partial ordering to minimize computation (\autoref{img:example}),
which is particularly important during exploration of large neighborhoods.
Our implementation (\autoref{sec:evaluation}) further reduces computation using batch processing.
Consequently,
the practical runtime complexity of the streaming variant of \autoref{algo:histogram} is approximately equivalent to that of the original 1-dimensional Weisfeiler-Lehman algorithm~\cite{shervashidze2011weisfeiler}:
using $R$-hop neighborhoods, the complexity is $O(R|E|)$.

\noindgras{Discounting Histogram Elements for Concept Drift.}
In APT scenarios,
the long duration of a potential attack suggests that
a good model must include the system's long-term behavior.
However, 
system behavior often changes over time,
which results in changes in the underlying statistical properties of the streaming provenance graph.
This phenomenon is referred to as concept drift~\cite{tsymbal2004problem}.

\unicorn accounts for such changes in system behavior
through the use of exponential weight decay~\cite{klinkenberg2004learning} on
histogram element counts to gradually forget outdated data.
It assigns weights that are inversely proportional to the age of the data~\cite{koychev2000gradual}.
For each element $h$ in the histogram $H$,
as a new data item $x_{t}$ (\ie a hashed label $l_{i}(v)$ from \autoref{algo:histogram} \autoref{algo:histogram:label}) streams in at time $t$,
instead of counting $H_{h}$ as:

{\scriptsize
\begin{align}
H_{h} = \sum_{t}\mathbbm{1}_{x_{t} = h} \nonumber
\end{align}
}

\unicorn discounts the influence of existing data according to their age:

{\scriptsize
\begin{align}
L_{h} = \sum_{t}w_{t}\mathbbm{1}_{x_{t} = h} \nonumber
\end{align}
}

where $\mathbbm{1}_{cond}$ is an indicator function that returns 1 when $cond$ is true and 0 otherwise,
and $w_t = e^{-\lambda\Delta t}$.
\unicorn increments $t$ monotonically with the number of edges added to the graph.
We use $L$ instead of $H$ to denote weighted histograms (\autoref{tab:notations}).

\noindgras{Applicability in Intrusion Detection Scenarios.}
The gradually forgetting approach helps \unicorn focus on the current dynamics of system execution
(\ie the most recent part of the provenance graph),
as well as any parts of the graph that are causally related to the current execution (based on the path length),
while maintaining fading ``memory'' of the past,
the rate of which is controlled by the weight decay factor $\lambda$.
Notably,
regardless of how temporally distant an event has occurred from the current state of system execution,
if it is causally related to a current object/activity,
that event and its surrounding neighborhood will contribute to the histogram without discount.
Therefore,
provenance graphs ensure that relevant contextual information remains in the analysis regardless of time.
In \autoref{sec:design:model}, we discuss how
\unicorn models a system's evolutionary behavior. 
 
\subsection{Generating Graph Sketches}
\label{sec:design:sketch}
The graph histogram is a simple vector space graph statistic that
describes system execution.
However,
unlike traditional histogram-based similarity analysis~\cite{baker1998distributional, chapelle1999support, yang2007evaluating},
\unicorn constantly updates the histogram as new edges arrive.
Measuring the similarity between \emph{streaming} histograms is challenging,
because the number of histogram elements is not known \emph{a priori} and changes continuously.
Moreover, we should measure the similarity
based on the underlying \emph{distribution} of graph features, 
instead of absolute counts. 
However,
most existing machine learning~\cite{hearst1998support, doersch2016tutorial} and data mining~\cite{fayyad1996advances, hand2007principles} techniques
applicable to modeling system behavior from graph histograms
require fixed-length numerical vectors.

One naive solution is to enumerate all possible histogram elements beforehand,
\eg through combinatorial enumeration and manual feature engineering.
However, given a potentially large number of vertex and edge labels,
exacerbated by the multi-label nature of the graph and a possibly large number of neighborhood iterations,
the resulting histogram will be sparse and thus problematic in terms of space and time complexity.
Alternatively, manual feature engineering might reduce the size of the histogram,
but it is time-consuming and inevitably requires arbitrary handling of unseen histogram elements.

We instead use locality sensitive hashing~\cite{indyk1998approximate},
also called similarity-preserving data sketching~\cite{wang2014hashing},
which is commonly used in classifying~\cite{aggarwal2010classification} and filtering~\cite{bachrach2009sketching} high-cardinality data streams.
\unicorn employs HistoSketch~\cite{yang2017histosketch},
an approach based on consistent weighted sampling~\cite{manasse2010consistent},
that efficiently maintains compact and fixed-size sketches for streaming histograms.
HistoSketch is a constant time algorithm, 
so it is fast enough to support real-time streaming analysis on rapidly
growing provenance graphs.
It also unbiasedly approximates histograms based on their \emph{normalized, generalized} min-max (Jaccard) similarity~\cite{wu2017consistent}.
Jaccard similarity~\cite{chum2008near} has been successfully applied in a variety of machine-learning-based real-world problems,
such as document analysis~\cite{shrivastava2014defense}
and malware classification~\cite{raff2017malware} and detection~\cite{drew2016polymorphic}.
We review HistoSketch in Appendix~\autoref{sec:app:histosketch}. 
For proof of correctness of the algorithm,
we refer interested readers to the original work by Yang \etal~\cite{yang2017histosketch}.

\subsection{Learning Evolutionary Models}
\label{sec:design:model}
Given graph sketches and a similarity metric,
clustering is a common data mining approach to identify outliers.
However, conventional clustering approaches fail to capture a system's
evolutionary behavior~\cite{akoglu2015graph}.
APT scenarios are sufficiently long term that failing to capture this
behavior leads to too many false positives~\cite{manzoor2016fast}.
\unicorn leverages its streaming capability to create evolutionary models
that capture normal changes in system behavior.
Crucially, it builds these evolutionary models during training, not during
deployment.
Systems that dynamically evolve models during deployment risk poisoning
their models during an APT's drawn out attack phase~\cite{manzoor2016fast}.

\unicorn creates a series of temporally-ordered sketches during training.
It then clusters this sketch sequence from a single server
using the well-known K-medoids algorithm~\cite{kaufman2009finding},
using the silhouette coefficient to determine the optimal value of
$K$~\cite{rousseeuw1987silhouettes}.
The clusters represent ``meta-states'' of system execution,
\eg startup, initialization, steady state behavior.
\unicorn then uses both the temporal ordering of the sketches in all clusters
and the statistics of each cluster (\eg diameter, medoid)
to produce a model of the system's evolution.
\autoref{algo:model} describes the construction of the evolutionary model.
Each sketch $S(t)$ belongs to a single cluster indexed $k$.
The evolution $E$ is an ordered list of cluster indices, whose order is determined by the temporally-ordered sketches $S(t)$.
\begin{algorithm}
	\scriptsize
	\algsetup{linenosize=\tiny}
	\SetAlgoLined
	\DontPrintSemicolon
	\SetKwInOut{Input}{Input}\SetKwInOut{Output}{Output}
	\SetKwFunction{BelongsTo}{BelongsTo}\SetKwFunction{Empty}{Empty}\SetKwFunction{Tail}{Tail}
	\Input{Sketches $S(t), t = 0, \cdots, T$ of a streaming provenance graph, ordered by time $t$}
	\Output{Evolution List $E$}
	$E$ = \{\}\;
	\For{$t \leftarrow 0$ \KwTo $T$}{
		$k$ = \BelongsTo{S(t)}\;\tcc*[r]{$k \in \{1, \cdots, K\}$}
		\lIf{\Empty{$E$} $\mid\mid$ \Tail{$E$} != $k$}{$E = E::k$}
	}
	\caption{Generating an Evolution Trace}\label{algo:model}
\end{algorithm}

For each training instance,
\unicorn creates a model that captures the changes of system execution states during its runtime.
Intuitively,
this is similar to an automaton~\cite{sekar2001fast, jafarian2011gray} that tracks the state of system execution.
The final model consists of many sub-models from all the provenance graphs in the training data.
With evolutionary modeling,
\unicorn learns system behavior at many points in time;
with the gradually forgetting scheme (\autoref{sec:design:histogram}), at \emph{any} point in time,
\unicorn is able to focus on the most relevant activities.
 
\subsection{Detecting Anomalies}
\label{sec:design:deploy}
During deployment,
anomaly detection follows the same streaming paradigm described in previous sections. 
\unicorn periodically creates graph sketches as the histogram evolves
from the streaming provenance graph.
Given a graph sketch, 
\unicorn compares the sketch to all the sub-models learned during modeling, 
fitting it to a cluster in each sub-model. 
\unicorn assumes that monitoring starts from system boot
and tracks system state transitions within each sub-model.
To be considered valid in any sub-model, 
a sketch must either fit into the current state or (one of) the next state(s)
(\ie in the cases where the sketch captures state transition in system execution);
otherwise, it is considered anomalous.
Thus, we detect two forms of anomalous behavior: sketches that do not fit
into existing clusters and invalid transitions between clusters.
  
\section{Implementation}
\label{sec:implementation}
We use the vertex-centric graph processing framework, GraphChi~\cite{kyrola2012graphchi} to implement \unicorn's graph processing algorithms in C++;
we implement its data parsing and modeling components in Python.

GraphChi~\cite{kyrola2012graphchi} is a disk-based system that
efficiently computes on large graphs with billions of edges on a single computer.
Using GraphChi,
\unicorn achieves efficient analysis performance without
needing to store the entire provenance graph in memory.
\unicorn relies on two important features of GraphChi:

\noindemph{1)}
GraphChi uses a
Parallel Sliding Windows (PSW) algorithm to split the graph into shards,
with approximately the same number of edges in each shard;
it computes on each shard in parallel.
The algorithm allows fast vertex and edge updates to disk with only a small number of non-sequential disk accesses.
This allows \unicorn to analyze the whole provenance graph
independent of memory constraints.

\noindemph{2)}
\unicorn leverages 
GraphChi's efficient computation on streaming graphs.
Per edge updates are efficient in their use of I/O, and
selective scheduling reduces computation.
\unicorn's guaranteed partial ordering (\autoref{sec:design:histogram})
minimizes the number of vertices it visits even when the
neighborhood hop parameter, $R$, is large.
Batching edge additions,
rather than processing one edge at a time, makes processing even faster.

Appendix~\autoref{sec:availability} provides details on obtaining and testing our open-source
implementation.
 
\section{Evaluation}
\label{sec:evaluation}
We analyzed approximately 1.5 TB of system monitoring data
containing approximately 2 billion OS-level provenance records from various tracing systems,
demonstrating the applicability of our approach.
Our evaluation addresses the following research questions:

\noindgras{Q1.} Can \unicorn accurately detect anomalies in long-running systems under APT attacks?
(\autoref{sec:evaluation:streamspot}, \autoref{sec:evaluation:darpa}, \autoref{sec:evaluation:supply})

\noindgras{Q2.} How important are the design decisions we made that are
tailored to the characteristics of APTs?
(\autoref{sec:evaluation:supply})

\noindgras{Q3.} Does \unicorn's gradually forgetting scheme
improve understanding of system behavior?
(\autoref{sec:evaluation:supply})

\noindgras{Q4.} How effective are \unicorn's evolutionary models
compared to existing clustering-based approaches that use static snapshots?
(\autoref{sec:evaluation:streamspot})

\noindgras{Q5.} Is \unicorn fast enough to perform
realtime monitoring and detection without falling behind?
(\autoref{sec:evaluation:system})

\noindgras{Q6.} What are \unicorn's memory and CPU utilization during system execution?
(\autoref{sec:evaluation:cpumem})

We compare \unicorn to StreamSpot,
a state-of-the-art anomaly detector that has shown promising results for APT attacks.
We show that multiple factors lead to \unicorn's higher detection accuracy;
both
its multi-hop graph exploration (Q2) and the evolutionary modeling scheme (Q4)
are better suited for provenance-based APT detection.

We then explore the efficacy of \unicorn with three real-life APT attack datasets,
all of which were captured during a red-team vs. blue-team adversarial engagement organized by U.S. DARPA
and are publicly available~\cite{darpa}.
We separate DARPA's datasets based on the underlying provenance capture systems
(\ie Cadets, ClearScope, and THEIA).
We show that \unicorn can detect system anomalies during real APT campaigns (Q1).

Lastly,
we create our own simulated APT supply-chain attack datasets (SC-1 and SC-2) using CamFlow in a controlled lab environment
to demonstrate that \unicorn is performant in terms of processing speed (Q5) and CPU and memory efficiency (Q6).
We show that it is capable of detecting simulated APT attacks using evolutionary modeling (Q4)
with diverse normal and background activities.
We further demonstrate that detection capability can be improved with contextualized graph exploration (Q2),
and that the gradually forgetting scheme improves detection accuracy,
because it helps better understand system behavior (Q3).

Properly evaluating and benchmarking security systems is difficult.
We adhere to the guidelines proposed by Kouwe \etal~\cite{van2018benchmarking} 
to the best of our ability.
For example,
the testbed that we designed to create the SC datasets ensures 
proper documentation of experiment specifications and easy reproduction of evaluation data and results.
We also use 5-fold cross-validation to provide a more accurate evaluation~\cite{liu2018host}.

\subsection{\unicorn vs. State-of-the-Art}
\label{sec:evaluation:streamspot}
StreamSpot~\cite{manzoor2016fast} is a clustering-based anomaly detection system that processes streaming heterogeneous graphs.
It extracts local graph features from single node/edge labels through breadth-first traversal on each graph node and vectorizes them for classification.
StreamSpot models only a single snapshot of every training graph
and dynamically maintains its clusters during test time by updating the parameters of the clusters.

\begin{table}[h]
	\centering
	\resizebox{\columnwidth}{!}{\begin{tabular}{l c r r r r}
			Experiment & Dataset & \# of Graphs & Avg. $\mid$V$\mid$ & Avg. $\mid$E$\mid$ & Preprocessed Data Size (GiB)\\
			\hline\hline
			\multirow{6}{4em}{StreamSpot} & YouTube & 100 & 8,292 & 113,229 & 0.3\\
			& Gmail & 100 & 6,827 & 37,382 & 0.1\\
			& Download & 100 & 8,831 & 310,814 & 1\\
			& VGame & 100 & 8,637 & 112,958 & 0.4\\
			& CNN & 100 & 8,990 & 294,903 & 0.9\\
			& Attack & 100 & 8,891 & 28,423 & 0.1\\
			\hline
		\end{tabular}}
	\caption{Characteristics of the StreamSpot dataset. The dataset is publicly available only in a preprocessed format.}
	\label{table:eval:data:streamspot}
\end{table}

\noindgras{Experimental Dataset.}
The StreamSpot dataset contains information flow graphs derived from six scenarios, five of which are benign~\cite{streamspotdata}.
Each scenario runs 100 times, producing 100 graphs for each.
Using the Linux \texttt{SystemTap} logging system~\cite{jacob2008systemtap},
the benign scenarios record system calls from normal browsing activities, such as watching YouTube videos and checking Gmail,
while the attack scenarios involve a drive-by download from a malicious URL that exploits a Flash vulnerability and gains root access to the visiting host.
The original scripts and the precise attack are unknown to us, 
however, the datasets are publicly available~\cite{streamspotdata}, and
we confirmed with the authors that the information flow graphs were constructed from all system calls on a machine from the start of a task until its termination.
\autoref{table:eval:data:streamspot} summarizes the dataset.

We use this dataset to fairly compare \unicorn with StreamSpot.
These nicely isolated scenarios may not represent today's typical workloads. 
However, they provide insight into how the different systems perform relative to each other. 
Furthermore, we might also interpret them as a proxy for the design pattern of today's microservice architecture~\cite{namiot2014micro}.
As we will see next,
\unicorn performs particularly well in such scenarios,
but it is also capable of accurately detecting anomalies in more diverse, heterogeneous computing environments (\autoref{sec:evaluation:darpa} and \autoref{sec:evaluation:supply}).
We discuss this point further in~\autoref{sec:limitations}.

\begin{table}[h]
	\resizebox{\columnwidth}{!}{\begin{tabular}{l | c c c c }
			Experiment & Precision & Recall & Accuracy & F-Score \\
			\hline\hline
			StreamSpot (baseline) & 0.74 & \cellcolor{gray!25}N/A & 0.66 & \cellcolor{gray!25}N/A\\
			\hline
			$R = 1$ & \cellcolor{yellow!25}0.51 & 1.0 & \cellcolor{yellow!25}0.60 & 0.68\\
			\hline
			$R = 3$ & \cellcolor{cyan!25}0.98 & 0.93 & \cellcolor{cyan!25}0.96 & 0.94\\
			\hline
		\end{tabular}
	}
	\caption{Comparison to StreamSpot on the StreamSpot dataset.
		We estimate StreamSpot's average accuracy and precision from the figure included in the paper~\cite{manzoor2016fast}, which does not
		report exact values. They did not report recall or F-score.}
	\label{table:eval:results:streamspot}
\end{table}

\begin{table}
	\resizebox{\columnwidth}{!}{\begin{tabular}{l | c c c c c c }
 		Experiment & \# of Test Graphs & \# of FPs ($R$ = 1) & \# of FPs ($R$ = 3) \\
 		\hline\hline
 		YouTube & 25 & \cellcolor{yellow!25}14 & \cellcolor{cyan!25}0\\
 		\hline
 		Gmail & 25 & \cellcolor{yellow!25}19 & \cellcolor{cyan!25}0\\
 		 \hline
 		Download & 25 & \cellcolor{yellow!25}25 & \cellcolor{cyan!25}2 \\
 		\hline
 		VGame & 25 & \cellcolor{yellow!25}20 & \cellcolor{cyan!25}0\\
 		\hline
 		CNN & 25 & \cellcolor{yellow!25}18 & \cellcolor{cyan!25}0\\
		\hline
		\end{tabular}
	}
	\caption{Decomposition of \unicorn's false positive results of the StreamSpot dataset.}
	\label{table:eval:streamspot}
\end{table}

\noindgras{Experimental Results.}
We compare \unicorn to StreamSpot, using StreamSpot's own dataset.
We configure \unicorn to use a sketch size $|S| = 2000$ and examine
with different neighborhood sizes,
$R = 1$ (equivalent to StreamSpot) and $R = 3$.
As shown in~\autoref{table:eval:results:streamspot},
\unicorn's ability to trivially consider larger neighborhoods ($R=3$)
produces significant precision/accuracy improvement.
The detailed precision results in \autoref{table:eval:streamspot}
further show that analyzing larger neighborhoods greatly reduces the
false positive rate.
This supports our hypothesis that contextual analysis is crucial.
\unicorn raises false positive alarms only on the Download dataset,
the most diverse dataset of StreamSpot's benign datasets (also manifested in its large average number of edges in \autoref{table:eval:data:streamspot}).
We discuss the importance of graph exploration in more depth in \autoref{sec:evaluation:supply}.

The following sections
evaluate \unicorn on various real-life and simulated APT attacks.
Unfortunately, we were unable to evaluate StreamSpot on these datasets,
because it could not handle the large number of edge types present
nor could it scale to the size of the graphs.
 
\subsection{DARPA TC Datasets}
\label{sec:evaluation:darpa}
Next, we demonstrate that \unicorn can effectively detect APTs
utilizing data from a variety of different provenance capture systems.

DARPA's Transparent Computing program focuses on developing technologies and prototype systems
to provide both forensic and detection of APTs.

\begin{table}[h]
	\centering
	\resizebox{\columnwidth}{!}{\begin{tabular}{l c r r r r}
			Experiment & Dataset & \# of Graphs & Avg. $\mid$V$\mid$ & Avg. $\mid$E$\mid$ & Raw Data Size (GiB)\\
			\hline\hline
			\multirow{2}{4em}{DARPA CADETS} & Benign & 66 & 59,983 & 4,811,836 & 271\\
			& Attack & 8 & 386,548 & 5,160,963 & 38\\
			\hline
			\hline
			\multirow{2}{4em}{DARPA ClearScope} & Benign & 43 & 2,309 & 4,199,309 & 441\\
			& Attack & 51 & 11,769 & 4,273,003 & 432\\
			\hline
			\hline
			\multirow{2}{4em}{DARPA THEIA} & Benign & 2 & 19,461 & 1,913,202 & 4\\
			& Attack & 25 & 275,822 & 4,073,621 & 85\\
			\hline
	\end{tabular}}
	\caption{Characteristics of graph datasets used in the DARPA experiments.}
	\label{table:eval:data:darpa}
\end{table}

\noindgras{Experimental Datasets.}
The DARPA datasets (\autoref{table:eval:data:darpa}) were collected from a network of hosts during the 2-week long third adversarial engagement of the DARPA Transparent Computing program.
The engagement involved various teams responsible for collecting audit data from different platforms (\eg Linux, Windows, BSD),
launching attacks during the engagement period,
and analyzing the data to detect attacks and perform forensic analysis.
The red team that carried out attacks also generated benign background activity (\eg web browsing),
thus allowing us to model normal system behavior.

The CADETS dataset was captured via the \textbf{C}ausal, \textbf{A}daptive, \textbf{D}istributed, and \textbf{E}fficient \textbf{T}racing \textbf{S}ystem (CADETS) on FreeBSD~\cite{cadets}.
ClearScope instruments the entire Android mobile software stack to capture provenance of the operations of mobile devices~\cite{clearscope}.
THEIA is a system for tagging and tracking multi-level host events~\cite{theia}; it instruments Ubuntu Linux machines during the engagement.

The experiment simulated
an enterprise setup~\cite{milajerdi2019holmes}
including
security-critical services such as a web server, an SSH server, an Email server,
and an SMB server (for shared file access).
The red team carried out various nation-state and common threats through the use of,
\eg a Firefox backdoor, a Nginx backdoor, and phishing emails.
Detailed descriptions of the attacks are available online~\cite{darpa}.

\begin{table}[h]
	\resizebox{\columnwidth}{!}{\begin{tabular}{l | c c c c }
			Experiment & Precision & Recall & Accuracy & F-Score \\
			\hline\hline
			DARPA CADETS & 0.98 & 1.0 & 0.99 & 0.99\\
			\hline
			DARPA ClearScope & 0.98 & 1.0 & 0.98 & 0.99\\
			\hline
			DARPA THEIA & 1.0 & 1.0 & 1.0 & 1.0\\
			\hline
		\end{tabular}
	}
	\caption{Experimental results of the DARPA datasets.}
	\label{table:eval:results:darpa}
\end{table}

\noindgras{Experimental Results.}
We partition each benign dataset into a training set (90\% of the graphs) and
a test dataset (10\% of the graphs).
We use the same sketch size ($|S| = 2000$) and neighborhood hop ($R = 3$) as in the StreamSpot experiment.
\autoref{table:eval:results:darpa} shows that \unicorn's analytics framework generalizes to different provenance capture systems
and various provenance graph structures.
\unicorn's high performance suggests that it can accurately detect anomalies in long-running systems of various platforms.
During the engagement,
the red team launched APT attacks using different attack vectors
and the attacks account for less than 0.001\% of the audit data volume~\cite{milajerdi2019holmes}.
\unicorn's anomaly-based detection mechanism identifies those attacks without prior attack knowledge,
even though they are embedded in an abundance of benign activity.

We note that some existing systems (Holmes~\cite{milajerdi2019holmes} and Poirot~\cite{milajerdi2019poirot})
also use the DARPA dataset for evaluation. 
Comparison between \unicorn and these systems is difficult, 
because they use a rule-based approach that requires \emph{a priori} expert knowledge to construct a model. 
\unicorn is fundamentally different, using an unsupervised learning model, requiring no expert input.
However, \unicorn's performance is comparable based on the number of detected attacks:
\unicorn detects all attacks on FreeBSD and Linux as do Holmes and Poirot.
We discuss the differences between these systems and \unicorn in greater
detail in \autoref{sec:rw}.
 
\subsection{Supply Chain Attack Scenarios}
\label{sec:evaluation:supply}
We designed two APT attack scenarios to run in a controlled lab environment.
These experiments evaluate the importance of graph analysis and evolutionary modeling (this section and \autoref{sec:evaluation:parameter})
to show that \unicorn is able to perform efficient, realtime monitoring (\autoref{sec:evaluation:system}).
Additionally, we use these experiments to understand how \unicorn performs when faced with attacks
that behave similarly to normal system workloads.
We carefully design benign and attack scenarios to achieve this goal. 
Previous experiments, conducted by others, do not guarantee similarity between benign and attack scenarios.

\begin{table}[h]
	\centering
	\resizebox{\columnwidth}{!}{\begin{tabular}{l c r r r r}
			Experiment & Dataset & \# of Graphs & Avg. $\mid$V$\mid$ & Avg. $\mid$E$\mid$ & Raw Data Size (GiB)\\
			\hline\hline
			\multirow{2}{4em}{SC-1} & Benign & 125 & 265,424 & 975,226 & 64\\
			& Attack & 25 & 257,156 & 957,968 & 12\\
			\hline
			\hline
			\multirow{2}{4em}{SC-2} & Benign & 125 & 238,338 & 911,153 & 59\\
			& Attack & 25 & 243,658 & 949,887 & 12 \\
			\hline
	\end{tabular}}
	\caption{Characteristics of the datasets used in the supply-chain APT attack experiments.}
	\label{table:eval:data:supply}
\end{table}

\noindgras{Experimental Datasets.}
We simulated two APT supply-chain attacks (SC-1 and SC-2) on a
Continuous Integration (CI) platform 
and used CamFlow (v0.5.0) to capture whole-system provenance,
including background activity, during both benign and attack scenarios.
For each scenario,
the experiment ran for three days.
To facilitate reproduction,
we leverage virtualization technology as our test harness and provide automated scripts 
for push-button replication of the experiments that generated the data (see Appendix~\autoref{sec:availability} for details).

To simulate APT attacks, 
we follow the typical cyber kill chain model that consists of roughly 7 nonexclusive phases,
\ie \emph{reconnaissance} (identify a target and explore its vulnerabilities),
\emph{weaponize} (design a backdoor and a penetration plan),
\emph{delivery} (deliver the weapon),
\emph{exploitation} (victim triggers the vulnerability),
\emph{installation} (install the backdoor or malware),
\emph{command and control (C\&C)} (give remote instructions to the victim),
and \emph{actions on objectives}~\cite{yadav2015technical}.

In the SC-1 experiment,
the attacker identified an enterprise CI server that routinely
\texttt{wget}s Debian packages from various repositories.
She discovered that the server runs GNU \texttt{wget} version 1.17,
which is vulnerable to arbitrary remote file upload
when the victim requests a malicious URL to a compromised server (CVE-2016-4971)~\cite{wgetexploit} (\emph{reconnaissance}).
The attacker embedded a common remote access trojan (RAT) into a Debian package
and compromised one of the repositories 
so that any request to download the legitimate package is redirected to the attacker's FTP server 
that hosts the RAT-embedded package (\emph{delivery}).
As the CI server downloaded (\emph{exploitation}) and installed (\emph{installation}) the package,
it also unknowingly installed the trojan software.
The RAT established a C\&C channel with the attacker,
creating a reverse TCP shell session on the CI server (\emph{C\&C}).
The attacker then modified the CI server configuration (\emph{actions on objectives}) to gain control of the CI deployment output.
The SC-2 experiment had a similar setup
but the attacker exploited a different vulnerability from GNU Bash version 4.3,
which allows remote attackers to execute arbitrary code via crafted trailing strings after function definitions in Bash scripts (CVE-2014-6271).
Both scenarios represent the \emph{supply-chain compromise} as the initial access mechanism 
to subvert a company's software distribution channel to spread malware~\cite{shadowhammer, ccleaner}.

In both experiments, we model normal behavior of the victim system (\ie the CI server).
\autoref{table:eval:data:supply} summaries the datasets.
\begin{table}
	\centering
	\resizebox{\columnwidth}{!}{\begin{tabular}{l | l l l l l }
			& Batch Size & Sketch Size & Hop Count & Decay Factor & Sketch Interval \\
			\hline\hline
			Baseline & 6,000 & 2,000 & 3 & 0.02 & 3,000\\
			\hline
		\end{tabular}
	}
	\caption{\unicorn configurations for supply-chain APT attack scenarios.}
	\label{table:eval:baseline}
\end{table}

\begin{table}[h]
	\resizebox{\columnwidth}{!}{\begin{tabular}{l | c c c c }
			Experiment & Precision & Recall & Accuracy & F-Score \\
			\hline\hline
			SC-1 & 0.85 & 0.96 & 0.90 & 0.90\\
			\hline
			SC-2 & 0.75 & 0.80 & 0.77 & 0.78\\
			\hline
		\end{tabular}
	}
	\caption{Experimental results of the supply-chain APT attack scenarios.}
	\label{table:eval:results:supply}
\end{table}

\noindgras{Experimental Results.}
We split 125 benign graphs randomly into 5 groups
to enable 5-fold cross validation.
After we use \unicorn to model normal behavior on the training set,
which consists of 100 benign graphs (\ie 4 groups),
we evaluate it on the remaining 25 benign graphs (\ie the $5^{th}$ group) for false positive validation.
We also evaluate the model on the 25 attack graphs for false negatives (\autoref{table:eval:data:supply}).
We repeat this procedure for each group and report the mean evaluation results.
\autoref{table:eval:baseline} summarizes the configuration for the experiments and
\autoref{table:eval:results:supply} shows the experimental results.

We see in~\autoref{table:eval:results:supply} that \unicorn is able to detect attacks with high accuracy
with only a small number of false alarms (as reflected in precision and recall).
We observe that \unicorn creates many clusters in the sub-models during modeling, and
the majority of true alarms originate from the first several clusters as \unicorn tracks system state transition.
This suggests that \unicorn's evolutionary model captures system behavior changes 	
and that it is able to detect attacks in their early stages, \ie the initial supply-chain access point.
This has important implications.
First,
traditional clustering approaches that
use static snapshots to build the initial model
generate a large number of false positives (\autoref{sec:evaluation:streamspot}).
These can easily overwhelm system administrators and cause ``threat fatigue'',
leading to alert dismissal.
In supply-chain attacks, the attackers can take advantage of this initial stage to break into an enterprise network.
\unicorn reduces false positives as its evolutionary models precisely but flexibly define normal system behavior.
We further note that dynamically adapting the model during runtime is also suboptimal in APT scenarios,
because once the attackers break into the network from the initial supply chain,
they can guide the model to slowly and gradually penetrate the network without the model raising an alarm.
Second, while many APTs abandon stealth in later attack stages, 
making detection easier, 
\unicorn raises alarms in earlier stages, 
thus preventing damage that may have already occurred when APTs unmask their behavior.
For example, we observe that
50\% of the attacks in SC-1 were detected when malicious packages were just delivered to the victim machine,
and all of them were flagged right after installation.

It is more difficult to detect attacks in the SC scenarios than in the DARPA ones.
The attackers in the DARPA datasets spend time finding vulnerabilities in the
system, and that behavior appears in the traces.
In contrast, in the SC scenarios, the attacker has \emph{a priori} knowledge of the
target system
(\ie we act as both the attacker and the victim), so we can launch an attack
without any prior unusual behavior.
This partially explains \unicorn's lower performance on the SC datasets.

In~\autoref{sec:evaluation:parameter}, we conduct additional experiments on SC-1 to demonstrate the importance of graph exploration
in detecting anomalies in provenance graphs.

\subsection{Influence of Graph Analysis on Detection Performance}
\label{sec:evaluation:parameter}
\definecolor{bblue}{HTML}{4F81BD}
\definecolor{rred}{HTML}{C0504D}
\definecolor{ggreen}{HTML}{9BBB59}
\definecolor{ppurple}{HTML}{9F4C7C}
\pgfplotsset{ /pgfplots/ybar legend/.style={ /pgfplots/legend image code/.code={ \draw[##1,/tikz/.cd,bar width=3pt,yshift=-0.2em,bar shift=0pt] plot coordinates {(2*\pgfplotbarwidth,0.6em)};}, } }
\pgfplotsset{compat=1.14}

\begin{figure*}
\renewcommand\thesubfigure{(\alph{subfigure})}
	\centering
	\caption[]{Detection performance (precision, recall, accuracy, and F-score) with varying hop counts (\autoref{fig:paramhop}), sketch sizes (\autoref{fig:paramsketch}), intervals of sketch generation (\autoref{fig:paraminterval}),
		and decay factor (\autoref{fig:paramdecay}).
		Baseline values (*) are used by the controlled parameters (that remain constant) in each figure.}
	\label{fig:perf:eval}
	\begin{tikzpicture}
	
	\begin{groupplot}[group style={
	    group size=2 by 2,
            xticklabels at=edge bottom,
            yticklabels at=edge left
        }, 
        height=0.3\textwidth,
        width=\columnwidth,
        major x tick style = transparent,
        ybar=2*\pgflinewidth,
        /pgf/bar width=5pt,
        ymin=0,
        xtick=data,
        ymajorgrids=true,
        scaled y ticks = false,
        legend to name=named,
        legend columns=-1
    ]
    
    \nextgroupplot[symbolic x coords={1,2,3*,4,5}, ylabel = {Rate}]
	\addplot[style={bblue,fill=bblue,mark=none}]
            coordinates {(1, 0.666666667) (2, 0.75) (3*, 0.892) (4, 0.566666667) (5, 0.6)};

        \addplot[style={rred,fill=rred,mark=none}]
             coordinates {(1, 0.569877345) (2, 0.628968254) (3*, 0.847078439) (4, 0.501298701) (5, 0.513131313)};

        \addplot[style={ggreen,fill=ggreen,mark=none}]
             coordinates {(1, 1) (2, 1) (3*, 0.96) (4, 0.8) (5, 1)};

        \addplot[style={ppurple,fill=ppurple,mark=none}]
             coordinates {(1, 0.721703297) (2, 0.771062271) (3*, 0.89946838) (4, 0.612820513) (5, 0.677380952)};
        
    \nextgroupplot[symbolic x coords={500,1000,2000*,3000,10000}]
    	\addplot[style={bblue,fill=bblue,mark=none}]
            coordinates {(500, 0.55) (1000, 0.633333333) (2000*, 0.892) (3000, 0.65) (10000, 0.666666667)};

        \addplot[style={rred,fill=rred,mark=none}]
             coordinates {(500, 0.40952381) (1000, 0.540909091) (2000*, 0.847078439) (3000, 0.544444444) (10000, 0.569877345)};

        \addplot[style={ggreen,fill=ggreen,mark=none}]
             coordinates {(500, 0.52) (1000, 1) (2000*, 0.96) (3000, 1) (10000, 1)};

        \addplot[style={ppurple,fill=ppurple,mark=none}]
             coordinates {(500, 0.438888889) (1000, 0.699358974) (2000*, 0.89946838) (3000, 0.704761905) (10000, 0.721703297)};
             
    \nextgroupplot[symbolic x coords={500,1000,3000*,5500}, ylabel = {Rate}]
    	 \addplot[style={bblue,fill=bblue,mark=none}]
            coordinates {(500, 0.583333333) (1000, 0.416666667) (3000*, 0.892) (5500, 0.466666667)};

        \addplot[style={rred,fill=rred,mark=none}]
             coordinates {(500, 0) (1000, 0.416666667) (3000*, 0.847078439) (5500, 0.439393939)};

        \addplot[style={ggreen,fill=ggreen,mark=none}]
             coordinates {(500, 0) (1000, 1) (3000*, 0.96) (5500, 1)};

        \addplot[style={ppurple,fill=ppurple,mark=none}]
             coordinates {(500, 0) (1000, 0.588235294) (3000*, 0.89946838) (5500, 0.610294118)};

    \nextgroupplot[symbolic x coords={0,0.02*,0.1,1}]
    	\addplot[style={bblue,fill=bblue,mark=none}]
            coordinates {(0, 0.583333333) (0.02*, 0.892) (0.1, 0.416666667) (1, 0.566666667)};

        \addplot[style={rred,fill=rred,mark=none}]
             coordinates {(0, 0) (0.02*, 0.847078439) (0.1, 0.416666667) (1, 0.25)};

        \addplot[style={ggreen,fill=ggreen,mark=none}]
             coordinates {(0, 0) (0.02*, 0.96) (0.1, 1) (1, 0.04)};

        \addplot[style={ppurple,fill=ppurple,mark=none}]
             coordinates {(0, 0) (0.02*, 0.89946838) (0.1, 0.588235294) (1, 0.285714286)};
  
    \legend{Accuracy, Precision, Recall, F-Score}
    \end{groupplot}
    \node[text width=6cm,align=center,anchor=north] at ([yshift=-3mm]group c1r1.south) {\captionof{subfigure}{Hop\label{fig:paramhop}}};
    \node[text width=6cm,align=center,anchor=north] at ([yshift=-3mm]group c2r1.south) {\captionof{subfigure}{Sketch\label{fig:paramsketch}}};
    \node[text width=6cm,align=center,anchor=north] at ([yshift=-3mm]group c1r2.south) {\captionof{subfigure}{Interval\label{fig:paraminterval}}};
    \node[text width=6cm,align=center,anchor=north] at ([yshift=-3mm]group c2r2.south) {\captionof{subfigure}{Decay\label{fig:paramdecay}}};
    \end{tikzpicture}
    \ref{named}
\end{figure*}

We now analyze the importance of \unicorn's key parameters using
the SC-1 dataset. 
We use the same setup from~\autoref{sec:evaluation:supply} as our baseline configurations (\autoref{table:eval:baseline}).
We then vary parameters independently to examine the impact of each.
\autoref{fig:perf:eval} shows the experimental results.

\noindent \emph{Batch Size (BS).}
This indicates the number of edges submitted to GraphChi at once;
it does not affect detection performance.

\noindent \emph{Hop Count (HC).} 
This defines the size of the neighborhood used to characterize each vertex;
it is a measure of the expressiveness of the features in our sketches.
Larger hop counts capture more contextual information, some of which might
be irrelevant, which can mask potential attacks~\cite{manzoor2016fast, milajerdi2019holmes},
as shown in \autoref{fig:paramhop}.

\noindent \emph{Sketch Size (SS).} 
This is the size of our fixed-size histogram representation.
Larger SS allows \unicorn to include more information about the evolving graph,
thus reducing the error of approximating normalized min-max similarity (\autoref{sec:design:sketch}).
However,
a large SS ultimately leads to the curse of dimensionality~\cite{friedman1997bias} in clustering (\autoref{sec:design:model}).
\autoref{fig:paramsketch} confirms that, in general, detection precision, recall, and accuracy improve
as we increase SS up to a point, after which it degrades.

\noindent \emph{Interval of Sketch Generation (SG).} 
This is the number of edges added to the graph between
the construction of new sketches.
Smaller SG makes adjacent sketches look similar to each other, which can
produce higher false negative rates and lower recall and accuracy.
Meanwhile, given fixed-size sketches,
which are approximations,
a larger SG leads to coarser-grained changes,
which also makes graphs look too similar to each other.
In \autoref{fig:paraminterval},
we observe that SG = $500$ edges per new sketch cannot detect any attack graphs,
resulting in 0 recall and undefined precision and F-score.
We obtain optimal results when setting the interval to be around $3,000$ in the SC-1 experiment.

\noindent \emph{Weighted Decay Factor (DF).} 
This determines the rate at which we forget the past.
We observe that both never-forget ($\lambda = 0.0$) and always-forget ($\lambda = 1.0$) yield unsatisfactory results,
while a slow decay rate (around $0.02$) achieves a good balance between
factoring past and current graph components into the analysis.
 
\subsection{Processing Speed}
\label{sec:evaluation:system}
In the previous section,
we show how \unicorn's parameters influence detection performance;
this section and the one that follows examine \unicorn's runtime overhead.
These sections consider only the CamFlow implementation,
which has been shown to produce negligible overhead~\cite{pasquier2017practical}, allowing us to isolate \unicorn's performance characteristics.
We use Amazon EC2 i3.2xlarge Linux machines with 8 vCPUs and 61GiB of memory.

Runtime performance is important in APT scenarios where an IDS is constantly monitoring the system in real time.
We analyze the SC-1 experiment
using the same baseline settings and parameters as in the previous section.
(We also evaluate batch size here, which has no impact on accuracy.)
Together, these two sections illustrate the tradeoff between accuracy and
runtime performance.

\begin{figure*}
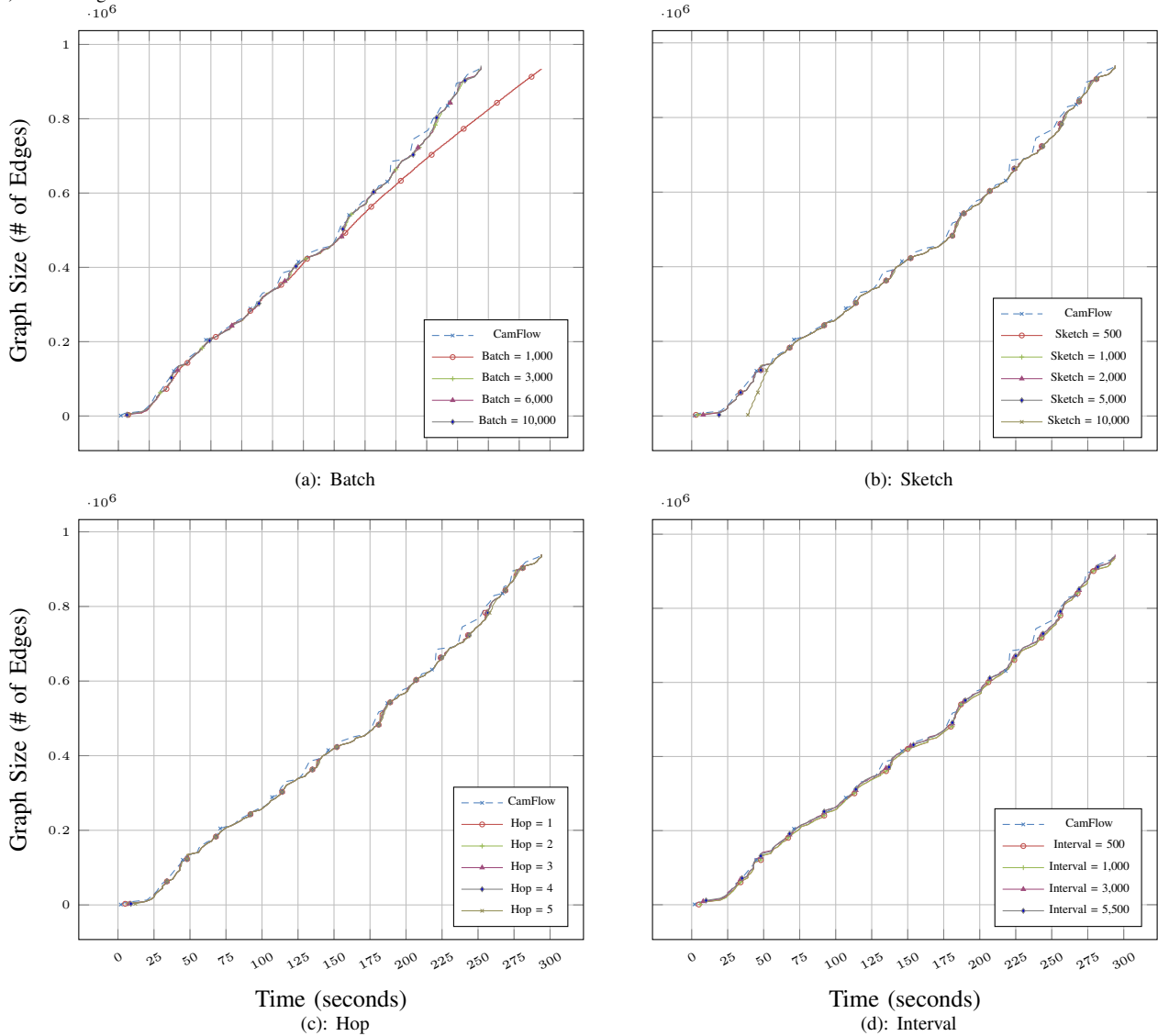

\renewcommand\thesubfigure{(\alph{subfigure})}
	\centering
	\caption[]{Total number of processed edges over time (in seconds) in the SC-1 experimental workload with varying batch sizes (\autoref{fig:perfbatch}),
		sketch sizes (\autoref{fig:perfsketch}), hop counts (\autoref{fig:perfhop}), and intervals of sketch generation (\autoref{fig:perfinterval}).
		Dashed blue line represents the speed of graph edges streamed into \unicorn for analysis.
		Triangle maroon baseline has the same configurations as those used in our experiments
		and indicates the values of the controlled parameters (that remain constant) in each figure.}
	\label{fig:perf}
	

\end{figure*}

\autoref{fig:perf} shows the total number of edges processed over time as a metric to quantify \unicorn's processing speed.
The CamFlow lines (in blue) represent the total number of edges generated by the capture system;
the closer other lines are to this line, the better the runtime performance,
meaning that \unicorn ``keeps up'' with the capture system.

\noindent \emph{Batch Size (BS).}
\autoref{fig:perfbatch} shows that runtime performance improves as we increase BS.
We use BS of 6,000 as it approximates CamFlow.
There is marginal improvement as we increase BS above 6,000.

\noindent \emph{Sketch Size (SS).}
As shown in~\autoref{fig:perfsketch},
SS has minimal impact on runtime performance,
except during the beginning of the experiment when \unicorn initializes the sketch,	
which requires more computation with larger sketch sizes.
Afterwards,
runtime performance is similar among different SS, 
thanks to \unicorn's fast, incremental sketch update.

\noindent \emph{Hop Count (HC).} 
\autoref{fig:perfhop} shows that HC has minimal impact on provenance graphs
because of the graph structure and its adaption of fast WL algorithm.

\noindent \emph{Interval of Sketch Generation (SG) \& Weighted Decay Factor (DF).} 
Neither SG nor DF affect runtime performance (we omit DF from the figure).

Overall, we show that \unicorn runtime is relatively insensitive to these parameters.
This means that \unicorn can perform realtime intrusion detection with
parameters optimized for detection accuracy.

\subsection{CPU \& Memory Utilization}
\label{sec:evaluation:cpumem}
We evaluate \unicorn's CPU utilization and memory overheads for a system under relatively heavy workload, \ie CI performing kernel compilation.
We show that \unicorn exhibits low CPU utilization and memory overheads.

\pgfplotstableread{
X   Y
1	0
2	0.625
3	7.625
4	9.875
5	11.15
6	11.85
7	12.2875
8	11.05
9	11.45
10	11.8
11	12.0125
12	12.25
13	11.275
14	10.9375
15	10.775
16	10.65
17	10.525
18	10.325
19	9.7125
20	9.65
21	9.4125
22	9.2875
23	9.1625
24	9.0875
25	8.6875
26	8.625
27	8.6125
28	8.6125
29	8.5875
30	9.175
31	8.3125
32	8.2875
33	8.3125
34	8.3375
35	8.35
36	8.1375
37	8.1375
38	8.175
39	8.175
40	8.225
41	8.225
42	8.075
43	8.1125
44	8.15
45	8.075
46	8.075
47	7.9125
48	7.8875
49	7.8875
50	7.875
51	7.925
52	8.0375
53	8.0125
54	8.1375
55	8.2375
56	8.3375
57	8.4375
58	8.525
59	8.4875
60	8.5625
61	8.6375
62	8.7125
63	8.7875
64	8.8875
65	8.825
66	8.8875
67	8.975
68	9.05
69	9.1125
70	9.0625
71	9.125
72	9.2125
73	9.2625
74	9.35
75	9.4
76	9.35
77	9.3875
78	9.475
79	9.5125
80	9.575
81	9.6125
82	9.6
83	9.6375
84	9.7125
85	9.725
86	9.8
87	9.825
88	9.8125
89	9.8375
90	9.8875
91	9.9125
92	9.975
93	9.9125
94	9.9625
95	10.0375
96	10.0375
97	10.0875
98	10.1375
99	10.075
100	10.125
101	10.125
102	10.175
103	10.1875
104	10.2375
105	10.175
106	10.225
107	10.2625
108	10.3
109	10.3125
110	10.35
111	10.325
112	10.3375
113	10.3625
114	10.4125
115	10.4375
116	10.3875
117	10.425
118	10.4625
119	10.4625
120	10.5125
121	10.5375
122	10.5125
123	10.5
124	10.55
125	10.575
126	10.625
127	10.65
128	10.575
129	10.625
130	10.65
131	10.6875
132	10.675
133	10.7125
134	10.6875
135	10.725
136	10.7375
137	10.7375
138	10.7625
139	10.75
140	10.7625
141	10.8
142	10.8125
143	10.825
144	10.85
145	10.825
146	10.8375
147	10.875
148	10.8875
149	10.8875
150	10.9125
151	10.8875
152	10.9125
153	10.95
154	10.9625
155	10.95
156	10.9875
157	10.95
158	10.9625
159	10.975
160	11
161	11.025
162	10.9875
163	11.0125
164	11
165	11.025
166	11.05
167	11.0625
168	11.0375
169	11.0625
170	11.0875
171	11.1
172	11.1
173	11.1125
174	11.0875
175	11.1125
176	11.125
177	11.15
178	11.175
179	11.1875
180	11.125
181	11.1375
182	11.1625
183	11.175
184	11.2
185	11.175
186	11.1875
187	11.2125
188	11.2125
189	11.2375
190	11.2375
191	11.2
192	11.225
193	11.2375
194	11.25
195	11.275
196	11.2875
197	11.275
198	11.275
199	11.2875
200	11.275
201	11.3
202	11.325
203	11.2875
204	11.3
205	11.325
206	11.3375
207	11.35
208	11.325
209	11.35
210	11.35
211	11.325
212	11.35
213	11.3625
214	11.35
215	11.35
216	11.35
217	11.375
218	11.3875
219	11.4125
220	11.375
221	11.3875
222	11.4
223	11.4125
224	11.425
225	11.4375
226	11.3875
227	11.3875
228	11.4
229	11.425
230	11.425
231	11.4
232	11.4125
233	11.4375
234	11.45
235	11.4625
236	11.4625
237	11.45
238	11.4625
239	11.475
240	11.4875
241	11.5
242	11.4875
243	11.45
244	11.4625
245	11.475
246	11.5
247	11.5125
248	11.5125
249	11.475
250	11.4875
251	11.5
252	11.525
253	11.5375
254	11.5
255	11.5125
256	11.525
257	11.5375
258	11.5375
259	11.5375
260	11.5125
261	11.5125
262	11.525
263	11.5375
264	11.55
265	11.5625
266	11.5375
267	11.55
268	11.55
269	11.575
270	11.5875
271	11.6
272	11.575
273	11.5875
274	11.6
275	11.6125
276	11.6
277	11.575
278	11.5875
279	11.6
280	11.5875
281	11.6
282	11.5875
283	11.575
284	11.5875
285	11.6
286	11.6
287	11.6125
288	11.625
289	11.6
290	11.6125
291	11.625
292	11.625
293	11.6375
294	11.65
295	11.6375
296	11.65
297	11.6375
298	11.65
299	11.6625
300	11.65
301	11.6625
302	11.6625
303	11.675
304	11.675
305	11.6625
306	11.625
307	11.6375
308	11.65
309	11.6625
310	11.675
311	11.6875
312	11.6625
313	11.675
314	11.6875
315	11.7
316	11.7
317	11.7
318	11.6875
319	11.7
320	11.7
321	11.7125
322	11.725
323	11.7
324	11.7125
325	11.725
326	11.7375
327	11.7375
328	11.7375
329	11.725
330	11.725
331	11.7375
332	11.75
333	11.7375
334	11.75
335	11.7125
336	11.7
337	11.7125
338	11.725
339	11.7375
340	11.7375
341	11.725
342	11.7375
343	11.7375
344	11.7375
345	11.75
346	11.725
347	11.7375
348	11.75
349	11.75
350	11.7625
351	11.775
352	11.75
353	11.7625
354	11.775
355	11.7875
356	11.775
357	11.7875
358	11.775
359	11.775
360	11.7875
361	11.8
362	11.8
363	11.8125
364	11.7875
365	11.8
366	11.8125
367	11.8125
368	11.825
369	11.775
370	11.7625
371	11.775
372	11.7875
373	11.7875
374	11.8
375	11.7875
376	11.7875
377	11.8
378	11.8125
379	11.8125
380	11.825
381	11.8
382	11.8125
383	11.8125
384	11.8125
385	11.825
386	11.8375
387	11.8125
388	11.825
389	11.825
390	11.8375
391	11.85
392	11.825
393	11.8375
394	11.85
395	11.85
396	11.8625
397	11.8625
398	11.8375
399	11.85
400	11.8625
401	11.8625
402	11.8625
403	11.875
404	11.85
405	11.85
406	11.8625
407	11.8625
408	11.875
409	11.8875
410	11.8625
411	11.875
412	11.875
413	11.875
414	11.875
415	11.8375
416	11.85
417	11.85
418	11.8625
419	11.8625
420	11.875
421	11.85
422	11.85
423	11.8625
424	11.8625
425	11.875
426	11.875
427	11.8625
428	11.875
429	11.875
430	11.8875
431	11.8875
432	11.9
433	11.8875
434	11.8875
435	11.9
436	11.9
437	11.9125
438	11.8875
439	11.8875
440	11.9
441	11.9
442	11.9125
443	11.9125
444	11.9
445	11.9
446	11.9125
447	11.9125
448	11.925
449	11.925
450	11.9125
451	11.925
452	11.925
453	11.9375
454	11.9375
455	11.95
456	11.925
457	11.9375
458	11.9375
459	11.95
460	11.9375
461	11.925
462	11.9375
463	11.9375
464	11.95
465	11.95
466	11.9625
467	11.9125
468	11.925
469	11.925
470	11.9375
471	11.925
472	11.925
473	11.9125
474	11.9125
475	11.925
476	11.925
477	11.925
478	11.9375
479	11.925
480	11.925
481	11.9375
482	11.9375
483	11.95
484	11.925
485	11.9375
486	11.9375
487	11.95
488	11.95
489	11.9625
490	11.9375
491	11.9375
492	11.9375
493	11.95
494	11.95
495	11.9625
496	11.95
497	11.95
498	11.9625
499	11.95
500	11.9625
501	11.9625
502	11.95
503	11.95
504	11.9625
505	11.9625
506	11.975
507	11.95
508	11.95
509	11.9625
510	11.9625
511	11.975
512	11.975
513	11.9625
514	11.95
515	11.9625
516	11.975
517	11.975
518	11.975
519	11.9625
520	11.975
521	11.975
522	11.975
523	11.975
524	11.975
525	11.9625
526	11.975
527	11.9875
528	11.9875
529	11.9875
530	11.975
531	11.975
532	11.975
533	11.975
534	11.9875
535	11.9875
536	11.975
537	11.9875
538	11.9875
539	12
540	11.9875
541	11.9875
542	11.975
543	11.9875
544	12
545	11.9875
546	12
547	12
548	11.9875
549	11.9875
550	12
551	11.9875
552	11.9875
553	11.9625
554	11.95
555	11.95
556	11.95
557	11.95
558	11.9625
559	11.95
560	11.95
561	11.95
562	11.9625
563	11.9625
564	11.975
565	11.9625
566	11.9625
567	11.9625
568	11.975
569	11.975
570	11.9875
571	11.975
572	11.975
573	11.975
574	11.9875
575	11.9875
576	11.975
577	11.9875
578	11.9875
579	11.9875
580	11.9875
581	11.9875
582	11.975
583	11.9875
584	11.9875
585	11.9875
586	12
587	12
588	11.9875
589	11.9875
590	12
591	12
592	12.0125
593	12.0125
594	12
595	12
596	12
597	12
598	12.0125
599	12
600	12
601	12
602	12.0125
603	12.0125
604	12.0125
605	12.0125
606	12.0125
607	12.0125
608	12.025
609	12.025
610	12.025
611	12.0125
612	12.025
613	12.025
614	12.025
615	12.025
616	12.025
617	12.0125
618	12.025
619	12.025
620	12.025
621	12.0375
622	12.025
623	12.025
624	12.025
625	12.0375
626	12.0375
627	12.0375
628	12.025
629	12.025
630	12.025
631	12.025
632	12.025
633	12.025
634	12.0125
635	12.025
636	12.0125
637	12.0125
638	12.025
639	12.025
640	12.0125
641	12.0125
642	12.025
643	12.025
644	12.0125
645	12.0125
646	12.0125
647	12.0125
648	12.025
649	12.025
650	12.025
651	12.0125
652	12.025
653	12.025
654	12.025
655	12.0375
656	12.0375
657	12.025
658	12.025
659	12.0375
660	12.0375
661	12.0375
662	12.05
663	12.025
664	12.025
665	12.0375
666	12.0375
667	12.0375
668	12.025
669	12.0375
670	12.0375
671	12.0375
672	12.05
673	12.05
674	12.0375
675	12.05
676	12.05
677	12.05
678	12.05
679	12.0625
680	12.05
681	12.05
682	12.0625
683	12.0625
684	12.0625
685	12.0625
686	12.05
687	12.05
688	12.05
689	12.0625
690	12.0625
691	12.05
692	12.05
693	12.0625
694	12.0625
695	12.0625
696	12.0625
697	12.0625
698	12.0625
699	12.0625
700	12.075
701	12.075
702	12.075
703	12.0625
704	12.075
705	12.075
706	12.075
707	12.075
708	12.0875
709	12.075
710	12.075
711	12.0875
712	12.0875
713	12.0875
714	12.075
715	12.0875
716	12.0875
717	12.0875
718	12.0875
719	12.1
720	12.0875
721	12.0875
722	12.0875
723	12.0875
724	12.0875
725	12.1
726	12.0875
727	12.0875
728	12.0875
729	12.1
730	12.1
731	12.1
732	12.075
733	12.0875
734	12.0875
735	12.0875
736	12.1
737	12.0875
738	12.0875
739	12.0875
740	12.1
741	12.1
742	12.1
743	12.0875
744	12.1
745	12.1
746	12.0875
747	12.0875
748	12.1
749	12.075
750	12.075
751	12.0875
752	12.0875
753	12.075
754	12.075
755	12.075
756	12.075
757	12.075
758	12.0875
759	12.075
760	12.0625
761	12.0625
762	12.075
763	12.075
764	12.075
765	12.0875
766	12.075
767	12.075
768	12.075
769	12.0875
770	12.0875
771	12.0875
772	12.075
773	12.0875
774	12.0875
775	12.0875
776	12.0875
777	12.1
778	12.0875
779	12.0875
780	12.0875
781	12.1
782	12.1
783	12.075
784	12.075
785	12.0875
786	12.0875
787	12.0875
788	12.0875
789	12.0875
790	12.0875
791	12.1
792	12.1
793	12.0875
794	12.0875
795	12.0875
796	12.0875
797	12.0875
798	12.0875
799	12.1
800	12.1
801	12.0875
802	12.0875
803	12.1
804	12.1
805	12.1
806	12.0875
807	12.1
808	12.1
809	12.1
810	12.1
811	12.1125
812	12.1
813	12.1
814	12.1
815	12.1125
816	12.1125
817	12.1125
818	12.1
819	12.1125
820	12.1125
821	12.1125
822	12.1125
823	12.125
824	12.1125
825	12.1125
826	12.1125
827	12.125
828	12.125
829	12.1125
830	12.1125
831	12.125
832	12.125
833	12.125
834	12.125
835	12.125
836	12.125
837	12.125
838	12.125
839	12.1375
840	12.125
841	12.1125
842	12.1125
843	12.125
844	12.125
845	12.125
846	12.125
847	12.125
848	12.125
849	12.125
850	12.125
851	12.1375
852	12.125
853	12.125
854	12.125
855	12.1375
856	12.1375
857	12.125
858	12.125
859	12.125
860	12.125
861	12.125
862	12.1375
863	12.1375
864	12.125
865	12.125
866	12.1375
867	12.1375
868	12.1375
869	12.1375
870	12.1375
871	12.1375
872	12.1375
873	12.1375
874	12.15
875	12.1375
876	12.1375
877	12.1375
878	12.15
879	12.15
880	12.15
881	12.1375
882	12.15
883	12.15
884	12.15
885	12.15
886	12.15
887	12.15
888	12.15
889	12.15
890	12.15
891	12.1625
892	12.1625
893	12.15
894	12.15
895	12.1625
896	12.1625
897	12.1625
898	12.15
899	12.1625
900	12.1625
901	12.1625
902	12.1625
903	12.1625
904	12.1625
905	12.15
906	12.15
907	12.15
908	12.1625
909	12.1625
910	12.15
911	12.15
912	12.1625
913	12.1625
914	12.1625
915	12.1625
916	12.1625
917	12.1625
918	12.1625
919	12.15
920	12.1625
921	12.1375
922	12.1375
923	12.1375
924	12.1375
925	12.15
926	12.15
927	12.125
928	12.125
929	12.1375
930	12.1375
931	12.15
932	12.15
933	12.1375
934	12.125
935	12.1375
936	12.1375
937	12.1375
938	12.1375
939	12.1375
940	12.1375
941	12.1375
942	12.1375
943	12.1375
944	12.1375
945	12.1375
946	12.1375
947	12.1375
948	12.15
949	12.15
950	12.1375
951	12.1375
952	12.15
953	12.15
954	12.15
955	12.15
956	12.1375
957	12.15
958	12.15
959	12.15
960	12.15
961	12.1625
962	12.15
963	12.15
964	12.15
965	12.15
966	12.1625
967	12.15
968	12.15
969	12.15
970	12.1625
971	12.1625
972	12.1625
973	12.15
974	12.1625
975	12.1625
976	12.1625
977	12.1625
978	12.1625
979	12.15
980	12.15
981	12.15
982	12.15
983	12.1625
984	12.1625
985	12.15
986	12.15
987	12.1625
988	12.1625
989	12.1625
990	12.15
991	12.1625
992	12.1625
993	12.1625
994	12.1625
995	12.1625
996	12.15
997	12.15
998	12.1625
999	12.1625
1000	12.1625
1001	12.1625
1002	12.15
1003	12.1625
1004	12.1625
1005	12.1625
1006	12.1625
1007	12.1625
1008	12.1625
1009	12.1625
1010	12.1625
1011	12.1625
1012	12.175
1013	12.1625
1014	12.1625
1015	12.1625
1016	12.175
1017	12.175
1018	12.175
1019	12.1625
1020	12.175
1021	12.175
1022	12.175
1023	12.175
1024	12.175
1025	12.175
1026	12.175
1027	12.175
1028	12.175
1029	12.175
1030	12.1875
1031	12.175
1032	12.175
1033	12.175
1034	12.1875
1035	12.1875
1036	12.175
1037	12.175
1038	12.1875
1039	12.1875
1040	12.1875
1041	12.1875
1042	12.175
1043	12.1875
1044	12.1875
1045	12.1875
1046	12.1875
1047	12.1875
1048	12.1875
1049	12.1875
1050	12.1875
1051	12.1875
1052	12.2
1053	12.2
1054	12.1875
1055	12.1875
1056	12.2
1057	12.2
1058	12.2
1059	12.1875
1060	12.1875
1061	12.1875
1062	12.2
1063	12.2
1064	12.2
1065	12.1875
1066	12.1875
1067	12.1875
1068	12.2
1069	12.2
1070	12.2
1071	12.1875
1072	12.2
1073	12.2
1074	12.2
1075	12.2
1076	12.2
1077	12.1875
1078	12.1875
1079	12.1875
1080	12.1875
1081	12.1875
1082	12.1875
1083	12.1875
1084	12.1875
1085	12.1875
1086	12.2
1087	12.2
1088	12.1875
1089	12.1875
1090	12.2
1091	12.2
1092	12.2
1093	12.2
1094	12.1875
1095	12.2
1096	12.2
1097	12.2
1098	12.2
1099	12.2
1100	12.2
1101	12.2
1102	12.2
1103	12.2
1104	12.2
1105	12.2
1106	12.2
1107	12.2
1108	12.2
1109	12.2125
1110	12.2125
1111	12.2
1112	12.2
1113	12.2125
1114	12.2125
1115	12.2125
1116	12.2125
1117	12.2
1118	12.2125
1119	12.2125
1120	12.2125
1121	12.2125
1122	12.2125
1123	12.2
1124	12.2125
1125	12.2125
1126	12.2125
1127	12.2125
1128	12.2
1129	12.2125
1130	12.2125
1131	12.2125
1132	12.2125
1133	12.2125
1134	12.2125
1135	12.2125
1136	12.2125
1137	12.2125
1138	12.2125
1139	12.225
1140	12.2125
1141	12.2125
1142	12.2125
1143	12.225
1144	12.225
1145	12.225
1146	12.2125
1147	12.2125
1148	12.225
1149	12.225
1150	12.225
1151	12.2125
1152	12.225
1153	12.225
1154	12.225
1155	12.225
1156	12.225
1157	12.225
1158	12.225
1159	12.225
1160	12.225
1161	12.225
1162	12.2375
1163	12.225
1164	12.225
1165	12.225
1166	12.225
1167	12.2375
1168	12.2375
1169	12.225
1170	12.225
1171	12.2375
1172	12.2375
1173	12.2375
1174	12.2125
1175	12.225
1176	12.225
1177	12.225
1178	12.225
1179	12.225
1180	12.225
1181	12.225
1182	12.225
1183	12.225
1184	12.2125
1185	12.2125
1186	12.2125
1187	12.2125
1188	12.225
1189	12.2375
1190	12.2375
1191	12.2375
1192	12.225
1193	12.225
1194	12.225
1195	12.225
1196	12.225
1197	12.2125
1198	12.2125
1199	12.2
1200	12.2
1201	12.2
1202	12.2125
1203	12.2
1204	12.2
1205	12.1875
1206	12.2
1207	12.2
1208	12.2
1209	12.1875
1210	12.1875
1211	12.2
1212	12.2
1213	12.2
1214	12.2
1215	12.2
1216	12.2
1217	12.2
1218	12.2
1219	12.2
1220	12.2
1221	12.2
1222	12.2
1223	12.2
1224	12.2
1225	12.2
1226	12.2
1227	12.2
1228	12.2
1229	12.2
1230	12.2125
1231	12.2125
1232	12.2
1233	12.2
1234	12.2
1235	12.2125
1236	12.2125
1237	12.2125
1238	12.2
1239	12.2125
1240	12.2125
1241	12.2125
1242	12.2125
1243	12.2
1244	12.2125
1245	12.2125
1246	12.2125
1247	12.2125
1248	12.2125
1249	12.2125
1250	12.2125
1251	12.2125
1252	12.2125
1253	12.2125
1254	12.2125
1255	12.2125
1256	12.2125
1257	12.2125
1258	12.2125
1259	12.2125
1260	12.2125
1261	12.2125
1262	12.2125
1263	12.2125
1264	12.2125
1265	12.2125
1266	12.2125
1267	12.2125
1268	12.2125
1269	12.2125
1270	12.2125
1271	12.225
1272	12.2125
1273	12.2125
1274	12.2125
1275	12.225
1276	12.225
1277	12.225
1278	12.2125
1279	12.2125
1280	12.225
1281	12.225
1282	12.225
1283	12.225
1284	12.225
1285	12.225
1286	12.225
1287	12.225
1288	12.225
1289	12.225
1290	12.225
1291	12.225
1292	12.225
1293	12.225
1294	12.225
1295	12.225
1296	12.225
1297	12.225
1298	12.225
1299	12.225
1300	12.2375
1301	12.225
1302	12.225
1303	12.225
1304	12.2375
1305	12.2375
1306	12.2375
1307	12.225
1308	12.2125
1309	12.225
1310	12.225
1311	12.225
1312	12.2125
1313	12.225
1314	12.225
1315	12.225
1316	12.225
1317	12.225
1318	12.225
1319	12.225
1320	12.225
1321	12.2375
1322	12.2375
1323	12.2375
1324	12.225
1325	12.225
1326	12.225
1327	12.225
1328	12.225
1329	12.225
1330	12.225
1331	12.225
1332	12.225
1333	12.225
1334	12.225
1335	12.225
1336	12.225
1337	12.225
1338	12.225
1339	12.225
1340	12.225
1341	12.225
1342	12.225
1343	12.225
1344	12.225
1345	12.225
1346	12.2375
1347	12.225
1348	12.225
1349	12.225
1350	12.2375
1351	12.2375
1352	12.2375
1353	12.225
1354	12.225
1355	12.2375
1356	12.2375
1357	12.2375
1358	12.225
1359	12.2375
1360	12.2375
1361	12.2375
1362	12.2375
1363	12.2375
1364	12.2375
1365	12.2375
1366	12.2375
1367	12.2375
1368	12.2375
1369	12.2375
1370	12.2375
1371	12.2375
1372	12.2375
1373	12.2375
1374	12.2375
1375	12.25
1376	12.2375
1377	12.2375
1378	12.2375
1379	12.2375
1380	12.25
1381	12.2375
1382	12.2375
1383	12.2375
1384	12.25
1385	12.25
1386	12.25
1387	12.2375
1388	12.2375
1389	12.2375
1390	12.2375
1391	12.2375
1392	12.25
1393	12.2375
1394	12.2375
1395	12.2375
1396	12.25
1397	12.25
1398	12.25
1399	12.2375
1400	12.2375
1401	12.25
1402	12.25
1403	12.25
1404	12.2375
1405	12.25
1406	12.25
1407	12.25
1408	12.25
1409	12.25
1410	12.25
1411	12.25
1412	12.25
1413	12.25
1414	12.25
1415	12.25
1416	12.25
1417	12.25
1418	12.25
1419	12.25
1420	12.25
1421	12.25
1422	12.25
1423	12.25
1424	12.25
1425	12.25
1426	12.25
1427	12.25
1428	12.25
1429	12.25
1430	12.25
1431	12.2625
1432	12.2625
1433	12.25
1434	12.25
1435	12.25
1436	12.2625
1437	12.2625
1438	12.2625
1439	12.25
1440	12.2625
1441	12.2625
1442	12.2625
1443	12.2625
1444	12.2625
1445	12.2625
1446	12.2625
1447	12.2625
1448	12.2625
1449	12.2625
1450	12.2625
1451	12.2625
1452	12.2625
1453	12.2625
1454	12.2625
1455	12.2625
1456	12.2625
1457	12.2625
1458	12.2625
1459	12.2625
1460	12.2625
1461	12.2625
1462	12.2625
1463	12.2625
1464	12.2625
1465	12.2625
1466	12.2625
1467	12.2625
1468	12.2625
1469	12.2625
1470	12.2625
1471	12.2625
1472	12.2625
1473	12.2625
1474	12.2625
1475	12.2625
1476	12.2625
1477	12.2625
1478	12.275
1479	12.2625
1480	12.2625
1481	12.2625
1482	12.2625
1483	12.275
1484	12.2625
1485	12.25
1486	12.25
1487	12.2625
1488	12.2625
1489	12.2625
1490	12.2625
1491	12.25
1492	12.2625
1493	12.2625
1494	12.2625
1495	12.2625
1496	12.25
1497	12.2625
1498	12.2625
1499	12.2625
1500	12.2625
1501	12.2625
1502	12.2625
1503	12.2625
1504	12.2625
1505	12.2625
1506	12.2625
1507	12.2625
1508	12.2625
1509	12.2625
1510	12.2625
1511	12.2625
1512	12.2625
1513	12.2625
1514	12.2625
1515	12.2625
1516	12.2625
1517	12.2625
1518	12.2625
1519	12.2625
1520	12.2625
1521	12.2625
1522	12.2625
1523	12.275
1524	12.275
1525	12.2625
1526	12.2625
1527	12.2625
1528	12.2625
1529	12.2625
1530	12.2625
1531	12.2625
1532	12.2625
1533	12.2625
1534	12.2625
1535	12.275
1536	12.275
1537	12.2625
1538	12.2625
1539	12.2625
1540	12.275
1541	12.275
1542	12.2625
1543	12.2625
1544	12.275
1545	12.275
1546	12.275
1547	12.275
1548	12.2625
1549	12.275
1550	12.275
1551	12.275
1552	12.275
1553	12.275
1554	12.275
1555	12.275
1556	12.275
1557	12.275
1558	12.275
1559	12.275
1560	12.275
1561	12.275
1562	12.275
1563	12.275
1564	12.275
1565	12.275
1566	12.275
1567	12.275
1568	12.275
1569	12.275
1570	12.275
1571	12.275
1572	12.275
1573	12.275
1574	12.275
1575	12.275
1576	12.2875
1577	12.275
1578	12.275
1579	12.275
1580	12.275
1581	12.2875
1582	12.2875
1583	12.275
1584	12.275
1585	12.2875
1586	12.2875
1587	12.2875
1588	12.275
1589	12.275
1590	12.275
1591	12.275
1592	12.275
1593	12.2875
1594	12.275
1595	12.275
1596	12.275
1597	12.2875
1598	12.2875
1599	12.2875
1600	12.275
1601	12.275
1602	12.2875
1603	12.2875
1604	12.2875
1605	12.2875
1606	12.2875
1607	12.2875
1608	12.2875
1609	12.2875
1610	12.2875
1611	12.2875
1612	12.2875
1613	12.2875
1614	12.2875
1615	12.2875
1616	12.2875
1617	12.2875
1618	12.2875
1619	12.2875
1620	12.2875
1621	12.2875
1622	12.2875
1623	12.2875
1624	12.2875
1625	12.2875
1626	12.2875
1627	12.2875
1628	12.2875
1629	12.2875
1630	12.2875
1631	12.2875
1632	12.2875
1633	12.2875
1634	12.2875
1635	12.2875
1636	12.2875
1637	12.2875
1638	12.2875
1639	12.3
1640	12.2875
1641	12.2875
1642	12.2875
1643	12.3
1644	12.3
1645	12.3
1646	12.2875
1647	12.2875
1648	12.3
1649	12.3
1650	12.3
1651	12.3
1652	12.3
1653	12.3
1654	12.3
1655	12.3
1656	12.3
1657	12.2875
1658	12.2875
1659	12.2875
1660	12.3
1661	12.3
1662	12.3
1663	12.2875
1664	12.2875
1665	12.3
1666	12.3
1667	12.3
1668	12.3
1669	12.3
1670	12.3
1671	12.3
1672	12.3
1673	12.3
1674	12.3
1675	12.3
1676	12.3
1677	12.3
1678	12.3
1679	12.3
1680	12.3
1681	12.3
1682	12.3
1683	12.3
1684	12.3
1685	12.3
1686	12.3
1687	12.3
1688	12.3
1689	12.275
1690	12.275
1691	12.275
1692	12.2625
1693	12.2625
1694	12.2625
1695	12.2625
1696	12.25
1697	12.2625
1698	12.25
1699	12.25
1700	12.25
1701	12.25
1702	12.2625
1703	12.25
1704	12.25
1705	12.25
1706	12.25
1707	12.2625
1708	12.2625
1709	12.25
1710	12.25
1711	12.2625
1712	12.2625
1713	12.2625
1714	12.2625
1715	12.25
1716	12.2625
1717	12.2625
1718	12.2625
1719	12.2625
1720	12.2625
1721	12.2625
1722	12.25
1723	12.2625
1724	12.2625
1725	12.2625
1726	12.25
1727	12.25
1728	12.2625
1729	12.2625
1730	12.2625
1731	12.2625
1732	12.25
1733	12.2625
1734	12.2625
1735	12.2625
1736	12.2625
1737	12.2625
1738	12.2625
1739	12.2625
1740	12.2625
1741	12.2625
1742	12.2625
1743	12.2625
1744	12.2625
1745	12.2625
1746	12.2625
1747	12.2625
1748	12.2625
1749	12.2625
1750	12.2625
1751	12.2625
1752	12.2625
1753	12.2625
1754	12.2625
1755	12.2625
1756	12.2625
1757	12.2625
1758	12.2625
1759	12.2625
1760	12.2625
1761	12.2625
1762	12.2625
1763	12.2625
1764	12.2625
1765	12.2625
1766	12.275
1767	12.2625
1768	12.2625
1769	12.2625
1770	12.2625
1771	12.275
1772	12.2625
1773	12.2625
1774	12.2625
1775	12.275
1776	12.275
1777	12.275
1778	12.2625
1779	12.2625
1780	12.275
1781	12.275
1782	12.275
1783	12.275
1784	12.2625
1785	12.275
1786	12.275
1787	12.275
1788	12.275
1789	12.275
1790	12.2625
1791	12.2625
1792	12.275
1793	12.275
1794	12.275
1795	12.2625
1796	12.2625
1797	12.275
1798	12.275
1799	12.275
1800	12.275
1801	12.2625
1802	12.275
1803	12.275
1804	12.275
1805	12.275
1806	12.275
1807	12.275
1808	12.275
1809	12.275
1810	12.275
1811	12.275
1812	12.275
1813	12.275
1814	12.275
1815	12.275
1816	12.275
1817	12.275
1818	12.275
1819	12.275
1820	12.275
1821	12.275
1822	12.275
1823	12.275
1824	12.275
1825	12.275
1826	12.275
1827	12.275
1828	12.275
1829	12.275
1830	12.275
1831	12.275
1832	12.275
1833	12.275
1834	12.275
1835	12.2875
1836	12.275
1837	12.275
1838	12.275
1839	12.275
1840	12.2875
1841	12.275
1842	12.275
1843	12.275
1844	12.275
1845	12.2875
1846	12.2875
1847	12.275
1848	12.275
1849	12.275
1850	12.2875
1851	12.2875
1852	12.2875
1853	12.275
1854	12.2875
1855	12.275
1856	12.275
1857	12.2875
1858	12.2875
1859	12.275
1860	12.275
1861	12.275
1862	12.2875
1863	12.2875
1864	12.275
1865	12.275
1866	12.275
1867	12.2875
1868	12.2875
1869	12.2875
1870	12.275
1871	12.2875
1872	12.2875
1873	12.2875
1874	12.2875
1875	12.2875
1876	12.2875
1877	12.2875
1878	12.2875
1879	12.2875
1880	12.2875
1881	12.2875
1882	12.2875
1883	12.2875
1884	12.2875
1885	12.2875
1886	12.2875
1887	12.2875
1888	12.2875
1889	12.2875
1890	12.2875
1891	12.2875
1892	12.2875
1893	12.2875
1894	12.2875
1895	12.2875
1896	12.2875
1897	12.2875
1898	12.2875
1899	12.2875
1900	12.2875
1901	12.2875
1902	12.2875
1903	12.2875
1904	12.2875
1905	12.2875
1906	12.2875
1907	12.2875
1908	12.2875
1909	12.2875
1910	12.2875
1911	12.2875
1912	12.2875
1913	12.2875
1914	12.2875
1915	12.3
1916	12.2875
1917	12.2875
1918	12.2875
1919	12.2875
1920	12.3
1921	12.3
1922	12.2875
1923	12.2875
1924	12.2875
1925	12.3
1926	12.2875
1927	12.3
1928	12.2875
1929	12.2875
1930	12.2875
1931	12.2875
1932	12.3
1933	12.2875
1934	12.2875
1935	12.2875
1936	12.2875
1937	12.3
1938	12.3
1939	12.2875
1940	12.2875
1941	12.2875
1942	12.3
1943	12.3
1944	12.3
1945	12.2875
1946	12.2875
1947	12.3
1948	12.3
1949	12.3
1950	12.3
1951	12.3
1952	12.3
1953	12.3
1954	12.3
1955	12.3
1956	12.3
1957	12.3
1958	12.3
1959	12.3
1960	12.3
1961	12.3
1962	12.3
1963	12.3
1964	12.3
1965	12.3
1966	12.3
1967	12.3
1968	12.3
1969	12.3
1970	12.3
1971	12.3
1972	12.3
1973	12.3
1974	12.3
1975	12.3
1976	12.3
1977	12.2875
1978	12.2875
1979	12.2875
1980	12.2875
1981	12.2875
1982	12.2875
1983	12.2875
1984	12.2875
1985	12.2875
1986	12.2875
1987	12.2875
1988	12.2875
1989	12.3
1990	12.3
1991	12.3
1992	12.3
1993	12.3
1994	12.3
1995	12.3
1996	12.3
1997	12.3
1998	12.3
1999	12.3
2000	12.3
2001	12.3
2002	12.2875
2003	12.2875
2004	12.2875
2005	12.2875
2006	12.2875
2007	12.2875
2008	12.2875
2009	12.2875
2010	12.2875
2011	12.2875
2012	12.2875
2013	12.3
2014	12.2875
2015	12.2875
2016	12.2875
2017	12.2875
2018	12.3
2019	12.3
2020	12.2875
2021	12.2875
2022	12.3
2023	12.3
2024	12.3
2025	12.2875
2026	12.2875
2027	12.3
2028	12.3
2029	12.3
2030	12.3
2031	12.2875
2032	12.3
2033	12.3
2034	12.3
2035	12.3
2036	12.3
2037	12.3
2038	12.3
2039	12.3
2040	12.3
2041	12.3
2042	12.3
2043	12.3
2044	12.3
2045	12.3
2046	12.3
2047	12.3
2048	12.3
2049	12.3
2050	12.3
2051	12.3
2052	12.3
2053	12.3
2054	12.3
2055	12.3
2056	12.3
2057	12.3
2058	12.3
2059	12.3
2060	12.3
2061	12.3
2062	12.3
2063	12.3
2064	12.3
2065	12.3
2066	12.3
2067	12.3
2068	12.3
2069	12.3
2070	12.3
2071	12.3
2072	12.3
2073	12.3
2074	12.3
2075	12.3
2076	12.3
2077	12.3
2078	12.3
2079	12.3
2080	12.3
2081	12.3
2082	12.3
2083	12.3
2084	12.3
2085	12.3
2086	12.3
2087	12.3
2088	12.3
2089	12.3
2090	12.3
2091	12.3
2092	12.3
2093	12.3
2094	12.3
2095	12.3
2096	12.3
2097	12.3
2098	12.3
2099	12.3
2100	12.3
2101	12.3
2102	12.3
2103	12.3
2104	12.3125
2105	12.3125
2106	12.3
2107	12.3
2108	12.3
2109	12.3125
2110	12.3125
2111	12.3125
2112	12.3
2113	12.3
2114	12.3125
2115	12.3125
2116	12.3125
2117	12.3
2118	12.3
2119	12.3125
2120	12.3125
2121	12.3125
2122	12.3125
2123	12.3
2124	12.3125
2125	12.3
2126	12.3125
2127	12.3125
2128	12.3125
2129	12.3
2130	12.3
2131	12.3125
2132	12.3125
2133	12.3125
2134	12.3125
2135	12.3
2136	12.3125
2137	12.3125
2138	12.3125
2139	12.3125
2140	12.3
2141	12.3125
2142	12.3125
2143	12.3125
2144	12.3125
2145	12.3125
2146	12.3125
2147	12.3125
2148	12.3125
2149	12.3125
2150	12.3125
2151	12.3125
2152	12.3125
2153	12.3125
2154	12.3125
2155	12.3125
2156	12.3125
2157	12.3125
2158	12.3125
2159	12.3125
2160	12.3125
2161	12.3125
2162	12.3125
2163	12.3125
2164	12.3125
2165	12.3125
2166	12.3125
2167	12.3125
2168	12.3125
2169	12.3125
2170	12.3125
2171	12.3125
2172	12.3125
2173	12.3125
2174	12.3125
2175	12.3125
2176	12.3125
2177	12.3125
2178	12.3125
2179	12.3125
2180	12.3125
2181	12.3125
2182	12.3125
2183	12.3125
2184	12.3125
2185	12.3125
2186	12.3125
2187	12.3125
2188	12.3125
2189	12.3125
2190	12.3125
2191	12.3125
2192	12.3125
2193	12.3125
2194	12.3125
2195	12.3125
2196	12.3125
2197	12.3125
2198	12.3125
2199	12.3125
2200	12.3125
2201	12.3125
2202	12.3125
2203	12.3125
2204	12.3125
2205	12.3125
2206	12.3125
2207	12.3125
2208	12.3125
2209	12.3125
2210	12.3125
2211	12.3125
2212	12.3125
2213	12.3125
2214	12.325
2215	12.3125
2216	12.3125
2217	12.3125
2218	12.3125
2219	12.325
2220	12.325
2221	12.3125
2222	12.3125
2223	12.3125
2224	12.325
2225	12.325
2226	12.325
2227	12.3125
2228	12.3125
2229	12.325
2230	12.325
2231	12.325
2232	12.3125
2233	12.325
2234	12.325
2235	12.325
2236	12.325
2237	12.325
2238	12.325
2239	12.325
2240	12.325
2241	12.325
2242	12.325
2243	12.325
2244	12.325
2245	12.325
2246	12.325
2247	12.325
2248	12.325
2249	12.325
2250	12.325
2251	12.325
2252	12.325
2253	12.325
2254	12.325
2255	12.325
2256	12.325
2257	12.325
2258	12.325
2259	12.325
2260	12.325
2261	12.325
2262	12.325
2263	12.325
2264	12.325
2265	12.325
2266	12.325
2267	12.325
2268	12.325
2269	12.325
2270	12.325
2271	12.325
2272	12.325
2273	12.325
2274	12.325
2275	12.325
2276	12.325
2277	12.325
2278	12.325
2279	12.325
2280	12.325
2281	12.325
2282	12.325
2283	12.325
2284	12.325
2285	12.325
2286	12.325
2287	12.325
2288	12.325
2289	12.325
2290	12.325
2291	12.325
2292	12.325
2293	12.325
2294	12.325
2295	12.325
2296	12.325
2297	12.325
2298	12.325
2299	12.325
2300	12.325
2301	12.325
2302	12.325
2303	12.325
2304	12.325
2305	12.325
2306	12.325
2307	12.325
2308	12.325
2309	12.325
2310	12.325
2311	12.325
2312	12.325
2313	12.325
2314	12.325
2315	12.325
2316	12.325
2317	12.325
2318	12.325
2319	12.325
2320	12.325
2321	12.325
2322	12.325
2323	12.325
2324	12.325
2325	12.325
2326	12.325
2327	12.325
2328	12.325
2329	12.325
2330	12.325
2331	12.325
2332	12.325
2333	12.325
2334	12.325
2335	12.3375
2336	12.325
2337	12.325
2338	12.325
2339	12.325
2340	12.3375
2341	12.3375
2342	12.325
2343	12.325
2344	12.325
2345	12.3375
2346	12.3375
2347	12.3375
2348	12.325
2349	12.325
2350	12.3375
2351	12.3375
2352	12.3375
2353	12.325
2354	12.3375
2355	12.3375
2356	12.3375
2357	12.3375
2358	12.3375
2359	12.3375
2360	12.3375
2361	12.3375
2362	12.3375
2363	12.3375
2364	12.3375
2365	12.3375
2366	12.3375
2367	12.3375
2368	12.3375
2369	12.3375
2370	12.3375
2371	12.3375
2372	12.3375
2373	12.3375
2374	12.3375
2375	12.3375
2376	12.3375
2377	12.3375
2378	12.3375
2379	12.3375
2380	12.3375
2381	12.3375
2382	12.3375
2383	12.3375
2384	12.3375
2385	12.3375
2386	12.3375
2387	12.3375
2388	12.3375
2389	12.3375
2390	12.3375
2391	12.3375
2392	12.3375
2393	12.3375
2394	12.3375
2395	12.3375
2396	12.3375
2397	12.3375
2398	12.3375
2399	12.3375
2400	12.3375
2401	12.3375
2402	12.3375
2403	12.3375
2404	12.3375
2405	12.3375
2406	12.3375
2407	12.3375
2408	12.3375
2409	12.3375
2410	12.3375
2411	12.3375
2412	12.3375
2413	12.3375
2414	12.3375
2415	12.3375
2416	12.3375
2417	12.3375
2418	12.3375
2419	12.3375
2420	12.3375
2421	12.3375
2422	12.325
2423	12.325
2424	12.325
2425	12.325
2426	12.325
2427	12.325
2428	12.325
2429	12.325
2430	12.325
2431	12.325
2432	12.325
2433	12.325
2434	12.325
2435	12.325
2436	12.325
2437	12.325
2438	12.325
2439	12.325
2440	12.325
2441	12.325
2442	12.325
2443	12.325
2444	12.3375
2445	12.325
2446	12.325
2447	12.325
2448	12.325
2449	12.3375
2450	12.3375
2451	12.325
2452	12.325
2453	12.325
2454	12.3375
2455	12.3375
2456	12.3375
2457	12.325
2458	12.325
2459	12.325
2460	12.325
2461	12.325
2462	12.3375
2463	12.325
2464	12.325
2465	12.325
2466	12.325
2467	12.3375
2468	12.325
2469	12.325
2470	12.325
2471	12.325
2472	12.3375
2473	12.3375
2474	12.325
2475	12.325
2476	12.325
2477	12.3375
2478	12.3375
2479	12.3375
2480	12.325
2481	12.325
2482	12.3375
2483	12.3375
2484	12.3375
2485	12.3375
2486	12.325
2487	12.3375
2488	12.3375
2489	12.3375
2490	12.3375
2491	12.325
2492	12.3375
2493	12.3375
2494	12.3375
2495	12.3375
2496	12.3375
2497	12.3375
2498	12.3375
2499	12.3375
2500	12.3375
2501	12.3375
2502	12.3375
2503	12.3375
2504	12.3375
2505	12.3375
2506	12.3375
2507	12.3375
2508	12.3375
2509	12.3375
2510	12.3375
2511	12.3375
2512	12.3375
2513	12.3375
2514	12.3375
2515	12.3375
2516	12.3375
2517	12.3375
2518	12.3375
2519	12.3375
2520	12.3375
2521	12.3375
2522	12.3375
2523	12.3375
2524	12.3375
2525	12.3375
2526	12.3375
2527	12.3375
2528	12.3375
2529	12.3375
2530	12.3375
2531	12.3375
2532	12.3375
2533	12.3375
2534	12.3375
2535	12.3375
2536	12.3375
2537	12.3375
2538	12.3375
2539	12.3375
2540	12.3375
2541	12.3375
2542	12.3375
2543	12.3375
2544	12.3375
2545	12.3375
2546	12.3375
2547	12.3375
2548	12.3375
2549	12.3375
2550	12.3375
2551	12.3375
2552	12.3375
2553	12.3375
2554	12.3375
2555	12.3375
2556	12.3375
2557	12.3375
2558	12.3375
2559	12.3375
2560	12.3375
2561	12.3375
2562	12.3375
2563	12.3375
2564	12.3375
2565	12.3375
2566	12.3375
2567	12.3375
2568	12.3375
2569	12.3375
2570	12.3375
2571	12.3375
2572	12.3375
2573	12.3375
2574	12.3375
2575	12.3375
2576	12.3375
2577	12.3375
2578	12.3375
2579	12.3375
2580	12.3375
2581	12.3375
2582	12.3375
2583	12.3375
2584	12.3375
2585	12.3375
2586	12.3375
2587	12.3375
2588	12.35
2589	12.3375
2590	12.3375
2591	12.3375
2592	12.3375
2593	12.3375
2594	12.3375
2595	12.3375
2596	12.3375
2597	12.3375
2598	12.3375
2599	12.3375
2600	12.3375
2601	12.3375
2602	12.3375
2603	12.3375
2604	12.3375
2605	12.3375
2606	12.3375
2607	12.3375
2608	12.3375
2609	12.3375
2610	12.35
2611	12.35
2612	12.3375
2613	12.3375
2614	12.3375
2615	12.35
2616	12.35
2617	12.35
2618	12.3375
2619	12.3375
2620	12.35
2621	12.35
2622	12.35
2623	12.35
2624	12.3375
2625	12.35
2626	12.35
2627	12.35
2628	12.35
2629	12.3375
2630	12.35
2631	12.35
2632	12.35
2633	12.35
2634	12.35
2635	12.35
2636	12.35
2637	12.35
2638	12.35
2639	12.35
2640	12.35
2641	12.35
2642	12.35
2643	12.35
2644	12.35
2645	12.35
2646	12.35
2647	12.35
2648	12.35
2649	12.35
2650	12.35
2651	12.35
2652	12.35
2653	12.35
2654	12.35
2655	12.35
2656	12.35
2657	12.35
2658	12.35
2659	12.35
2660	12.35
2661	12.35
2662	12.35
2663	12.35
2664	12.35
2665	12.35
2666	12.35
2667	12.35
2668	12.35
2669	12.35
2670	12.35
2671	12.35
2672	12.35
2673	12.35
2674	12.35
2675	12.35
2676	12.35
2677	12.35
2678	12.35
2679	12.35
2680	12.35
2681	12.35
2682	12.35
2683	12.35
2684	12.35
2685	12.35
2686	12.35
2687	12.35
2688	12.35
2689	12.35
2690	12.35
2691	12.35
2692	12.35
2693	12.35
2694	12.35
2695	12.35
2696	12.35
2697	12.35
2698	12.35
2699	12.35
2700	12.35
2701	12.35
2702	12.35
2703	12.35
2704	12.35
2705	12.35
2706	12.35
2707	12.35
2708	12.35
2709	12.35
2710	12.35
2711	12.35
2712	12.35
2713	12.35
2714	12.35
2715	12.35
2716	12.35
2717	12.35
2718	12.35
2719	12.35
2720	12.35
2721	12.35
2722	12.35
2723	12.35
2724	12.35
2725	12.35
2726	12.35
2727	12.35
2728	12.35
2729	12.35
2730	12.35
2731	12.35
2732	12.35
2733	12.35
2734	12.35
2735	12.35
2736	12.35
2737	12.35
2738	12.35
2739	12.35
2740	12.35
2741	12.35
2742	12.35
2743	12.35
2744	12.35
2745	12.35
2746	12.35
2747	12.35
2748	12.35
2749	12.35
2750	12.35
2751	12.35
2752	12.35
2753	12.35
2754	12.35
2755	12.35
2756	12.35
2757	12.35
2758	12.35
2759	12.35
2760	12.3625
2761	12.3625
2762	12.35
2763	12.35
2764	12.35
2765	12.3625
2766	12.3625
2767	12.35
2768	12.35
2769	12.35
2770	12.3625
2771	12.3625
2772	12.3625
2773	12.35
2774	12.35
2775	12.3625
2776	12.3625
2777	12.3625
2778	12.3625
2779	12.35
2780	12.3625
2781	12.3625
2782	12.3625
2783	12.3625
2784	12.3625
2785	12.3625
2786	12.3625
2787	12.3625
2788	12.3625
2789	12.3625
2790	12.35
2791	12.35
2792	12.35
2793	12.3625
2794	12.3625
2795	12.3625
2796	12.35
2797	12.35
2798	12.3625
2799	12.3625
2800	12.3625
2801	12.3625
2802	12.35
2803	12.3625
2804	12.3625
2805	12.3625
2806	12.3625
2807	12.3625
2808	12.3625
2809	12.3625
2810	12.3625
2811	12.3625
2812	12.3625
2813	12.3625
2814	12.3625
2815	12.3625
2816	12.3625
2817	12.3625
2818	12.3625
2819	12.3625
2820	12.3625
2821	12.3625
2822	12.3625
2823	12.3625
2824	12.3625
2825	12.3625
2826	12.3625
2827	12.3625
2828	12.3625
2829	12.3625
2830	12.3625
2831	12.3625
2832	12.3625
2833	12.3625
2834	12.3625
2835	12.3625
2836	12.3625
2837	12.3625
2838	12.3625
2839	12.3625
2840	12.3625
2841	12.3625
2842	12.3625
2843	12.3625
2844	12.3625
2845	12.3625
2846	12.3625
2847	12.3625
2848	12.3625
2849	12.3625
2850	12.3625
2851	12.3625
2852	12.3625
2853	12.3625
2854	12.3625
2855	12.3625
2856	12.35
2857	12.35
2858	12.35
2859	12.35
2860	12.3625
2861	12.3625
2862	12.3625
2863	12.3625
2864	12.3625
2865	12.3625
2866	12.3625
2867	12.3625
2868	12.3625
2869	12.3625
2870	12.3625
2871	12.3625
2872	12.3625
2873	12.35
2874	12.35
2875	12.35
2876	12.35
2877	12.35
2878	12.35
2879	12.35
2880	12.35
2881	12.35
2882	12.35
2883	12.35
2884	12.35
2885	12.35
2886	12.35
2887	12.35
2888	12.35
2889	12.35
2890	12.35
2891	12.35
2892	12.35
2893	12.35
2894	12.3375
2895	12.3375
2896	12.3375
2897	12.3375
2898	12.3375
2899	12.3375
2900	12.3375
2901	12.3375
2902	12.3375
2903	12.325
2904	12.325
2905	12.325
2906	12.325
2907	12.325
2908	12.325
2909	12.325
2910	12.325
2911	12.325
2912	12.325
2913	12.325
2914	12.325
2915	12.325
2916	12.325
2917	12.325
2918	12.325
2919	12.325
2920	12.325
2921	12.325
2922	12.325
2923	12.325
2924	12.325
2925	12.325
2926	12.325
2927	12.3375
2928	12.325
2929	12.325
2930	12.325
2931	12.325
2932	12.325
2933	12.325
2934	12.325
2935	12.325
2936	12.325
2937	12.325
2938	12.325
2939	12.325
2940	12.325
2941	12.325
2942	12.325
2943	12.325
2944	12.325
2945	12.325
2946	12.325
2947	12.325
2948	12.325
2949	12.325
2950	12.3375
2951	12.325
2952	12.325
2953	12.325
2954	12.325
2955	12.3375
2956	12.3375
2957	12.325
2958	12.325
2959	12.325
2960	12.3375
2961	12.3375
2962	12.3375
2963	12.325
2964	12.325
2965	12.3375
2966	12.3375
2967	12.3375
2968	12.325
2969	12.325
2970	12.3375
2971	12.3375
2972	12.3375
2973	12.3375
2974	12.325
2975	12.3375
2976	12.3375
2977	12.3375
2978	12.3375
2979	12.3375
2980	12.3375
2981	12.3375
2982	12.3375
2983	12.3375
2984	12.3375
2985	12.3375
2986	12.3375
2987	12.3375
2988	12.3375
2989	12.3375
2990	12.3375
2991	12.3375
2992	12.3375
2993	12.3375
2994	12.3375
2995	12.3375
2996	12.3375
2997	12.325
2998	12.3375
2999	12.3375
3000	12.3375
3001	12.3375
3002	12.3375
3003	12.3375
3004	12.3375
3005	12.3375
3006	12.3375
3007	12.3375
3008	12.3375
3009	12.3375
3010	12.3375
3011	12.3375
3012	12.3375
3013	12.3375
3014	12.3375
3015	12.3375
3016	12.3375
3017	12.3375
3018	12.3375
3019	12.3375
3020	12.3375
3021	12.3375
3022	12.3375
3023	12.3375
3024	12.3375
3025	12.3375
3026	12.3375
3027	12.3375
3028	12.3375
3029	12.3375
3030	12.3375
3031	12.3375
3032	12.3375
3033	12.3375
3034	12.3375
3035	12.3375
3036	12.3375
3037	12.3375
3038	12.3375
3039	12.3375
3040	12.3375
3041	12.3375
3042	12.3375
3043	12.3375
3044	12.3375
3045	12.3375
3046	12.3375
3047	12.3375
3048	12.3375
3049	12.3375
3050	12.3375
3051	12.3375
3052	12.3375
3053	12.3375
3054	12.3375
3055	12.3375
3056	12.3375
3057	12.3375
3058	12.325
3059	12.325
3060	12.325
3061	12.325
3062	12.325
3063	12.325
3064	12.3375
3065	12.3375
3066	12.3375
3067	12.3375
3068	12.3375
3069	12.3375
3070	12.3375
3071	12.3375
3072	12.3375
3073	12.3375
3074	12.3375
3075	12.3375
3076	12.3375
3077	12.3375
3078	12.3375
3079	12.3375
3080	12.3375
3081	12.3375
3082	12.3375
3083	12.3375
3084	12.3375
3085	12.3375
3086	12.3375
3087	12.3375
3088	12.3375
3089	12.3375
3090	12.3375
3091	12.3375
3092	12.3375
3093	12.325
3094	12.325
3095	12.325
3096	12.325
3097	12.325
3098	12.325
3099	12.325
3100	12.325
3101	12.325
3102	12.325
3103	12.325
3104	12.325
3105	12.325
3106	12.325
3107	12.3375
3108	12.3375
3109	12.3375
3110	12.3375
3111	12.3375
3112	12.3375
3113	12.3375
3114	12.3375
3115	12.3375
3116	12.3375
3117	12.3375
3118	12.3375
3119	12.3375
3120	12.3375
3121	12.3375
3122	12.3375
3123	12.3375
3124	12.3375
3125	12.3375
3126	12.3375
3127	12.325
3128	12.325
3129	12.325
3130	12.325
3131	12.325
3132	12.325
3133	12.325
3134	12.325
3135	12.325
3136	12.325
3137	12.325
3138	12.325
3139	12.325
3140	12.325
3141	12.325
3142	12.325
3143	12.325
3144	12.325
3145	12.325
3146	12.325
3147	12.325
3148	12.325
3149	12.325
3150	12.325
3151	12.325
3152	12.325
3153	12.325
3154	12.325
3155	12.325
3156	12.325
}\longcputable

\definecolor{bblue}{HTML}{4F81BD}
\definecolor{rred}{HTML}{C0504D}
\definecolor{ggreen}{HTML}{9BBB59}
\definecolor{ppurple}{HTML}{9F4C7C}

\autoref{fig:cpu:long} shows the average CPU utilization due to \unicorn over a long-running experiment using the baseline configuration.
The average CPU utilization stabilizes around 12.3\% on a single CPU.
\autoref{fig:cpu:short} shows the per-vCPU and the average CPU utilization of the same experiment (but with a shorter timeline to avoid cluttering). 
Note that the parameters discussed in the previous sections do not
significantly impact average CPU utilization.

PassMark Software's Enterprise Endpoint Security performance benchmark~\cite{passmark}
shows an average CPU utilization of 26\%-90\% for commercial products such as Symantec Endpoint Protection,
Kaspersky Endpoint Security,
and McAfee Endpoint Security. 
However, direct comparison is difficult. 
Research IDS we surveyed (\autoref{sec:rw}) do not report metrics that allow meaningful comparison.
We leave the design of meaningful performance benchmarks for IDS to future work.

\begin{table}[t]
	\centering
	\resizebox{\columnwidth}{!}{\begin{tabular}{l l c}
			Configuration Parameter & Parameter Value & Max Memory Usage (MB)\\
			\hline\hline
			\multirow{5}{4em}{Hop Count} & R = 1 & 562\\
			& R = 2 & 624\\
			& \cellcolor{cyan!25}R = 3 & \cellcolor{cyan!25}687\\
			& R = 4 & 749\\
			& R = 5 & 812\\
			\hline
			\hline
			\multirow{5}{4em}{Sketch Size} 
			& $|S|$ = 500 & 312\\
			& $|S|$ = 1,000 & 437\\
			& \cellcolor{cyan!25}$|S|$ = 2,000 & \cellcolor{cyan!25}687\\
			& $|S|$ = 5,000 & 1,374\\
			& $|S|$ = 10,000 & 2,498\\
			\hline
		\end{tabular}
	}
	\caption{Memory overheads with varying hop counts and sketch sizes. The highlighted configurations gave the best detection performance.}
	\label{table:mem}
\end{table}

\autoref{table:mem} shows memory overheads for the same workload under two different parameters.
Other parameters in the basic configurations do not significantly influence memory consumption.

The graph histogram and the random variables sampled for sketch generation represent the majority of \unicorn's memory usage.
The size of the histogram is proportional to the number of unique labels,
which in turn is determined by the size of the neighborhood that each vertex explores.
\autoref{table:mem} shows that as we increase the number of neighborhood hops,
\unicorn requires more memory for the histogram.
In theory,
the total number of unique labels is bounded by
the number of possible combinations of node/edge types within $|R|$ hops.
In practice, however,
many parts of a provenance graph exhibit similar structures
and therefore the value is significantly lower than the theoretical upper bound.
For example,
the maximum memory usage of the experiment corresponding to \autoref{fig:cpu:long} remains around 680MB,
which is acceptable for modern systems.
As we increase the sketch size,
\unicorn consumes more memory to store additional random variables sampled for sketch generation and update.
Although with $|S| = 10,000$, memory consumption increases up to 2.5GiB (\autoref{table:mem}), 
the previous sections suggest that such large values are never a good option.

\begin{figure}[h!]
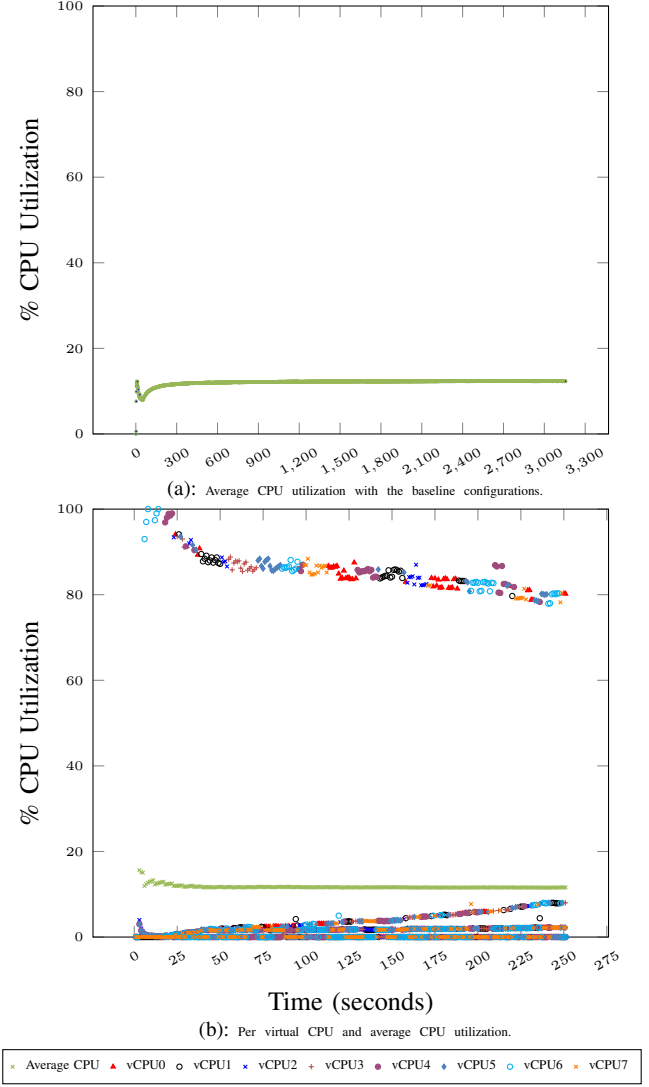

\renewcommand\thesubfigure{(\alph{subfigure})}
	\centering
	\caption{Evaluation of \unicorn's CPU utilization.}
	\label{fig:cpu}
	

        \ref{named}
\end{figure}  
\section{Discussion \& Limitations}
\label{sec:limitations}
\noindgras{Anomaly-based Detection.}
\unicorn shares assumptions, architecture, and therefore
limitations with other anomaly-based intrusion detection systems.

First,
we assume there is a modeling period where system administrators can run their
systems safely to capture system behavior (\autoref{sec:threat}).
Second,
we assume that there exists an exhaustive, finite number of
system behavior patterns and most,
if not all, patterns are observed during modeling~\cite{sekar2001fast}.
\unicorn will raise a false alarm if it observes a new
behavior that is, in fact, normal; such alarms will
require human-in-the-loop intervention.

Adversaries may try to steer malicious behavior towards learned models to evade detection,
similar to the well-known mimicry attacks~\cite{parampalli2008practical}.
However, conducting mimicry attacks on provenance graphs and/or \unicorn's graph sketches 
is more challenging than on sequences of system calls,
because provenance graphs contain complex structural information
that is difficult to imitate without affecting the attack.
Additionally, 
\unicorn's consistent weighted sampling approach randomizes sketch generation,
making it hard to guarantee that the low-dimensional projections of mimicry provenance graphs 
will be close to learned normal clusters.
At the same time,
this complexity will make it
difficult to use models to identify the cause of an alarm.
Fortunately,
there exist tools that facilitate attack causality analysis on provenance graphs (\autoref{sec:rw}).

\unicorn, like other anomaly-based systems, requires sufficient benign behavior traces
to learn behavior models.
Our current modeling design and implementation assumes that
\unicorn monitors a system from the same starting point as its model,
tracking state transitions as the system executes.
It can easily perform such continuous monitoring given its high performance and scalability (\autoref{sec:evaluation}).
However,
\unicorn may raise false alarms if the current state does not match the modeled state
as a result of, \eg the system restoring to a saved state due to failure.
One approach to addressing this issue is to integrate \unicorn 
more closely with the system so that it can save its model state at the same
time the system creates snapshots; when the system restores a snapshot,
\unicorn would restore the corresponding model state.

\noindgras{False Alarms.}
As briefly mentioned above,
when normal system behavior changes,
\unicorn might raise false positive alarms, since it does not dynamically
adjust its model (to avoid attacker poisoning).
The false alert problem is not unique to \unicorn.
In fact, it is a major concern even for signature-based approaches
used in practice, although in theory,
these approaches are designed to generate few such alarms due to rigid
attack matching~\cite{fireeye}.
\unicorn partially mitigates this issue using concept drift (\autoref{sec:design:histogram}),
modeling system evolution.
As we demonstrated in~\autoref{sec:evaluation},
\unicorn improves precision by 24\% compared to the state-of-the-art
and achieves near perfect results on the DARPA datasets.
However, \unicorn must also ensure stability to avoid adversarial manipulation,
which inevitably increases the potential for false alerts.
Therefore,
system administrators might need to periodically retrain \unicorn's
model to stay up-to-date.
Fortunately, updating \unicorn models is quick;
the important caveat is to ensure that
new training data is known to be devoid of malicious activity.

\noindgras{Graph Analysis.} 
\unicorn enables efficient and powerful graph analysis,
but similar to many IDS~\cite{manzoor2016fast, milajerdi2019holmes, chandola2009anomaly, akoglu2015graph},
it also requires parameters that must be tuned to each system to improve detection performance (\autoref{sec:evaluation:parameter}).
Finding ideal parameters for a specific task is a well-known research topic in machine learning~\cite{thornton2013auto}.
We used OpenTuner~\cite{ansel2014opentuner}
to enable program autotuning.
Although our tests are by no means exhaustive, 
it is encouraging that we were able to use the same settings on almost all of our evaluation datasets, 
even though they came from disparate sources and modeled rather different attacks.
Due to time and space constraints,
we leave the discussion of automatically tuning \unicorn to future work.

\noindgras{Heterogenous Host Activity.}
In testing, we observed that \unicorn performed extremely well in domains
featuring homogenous normal activities.
For example,
we achieve near-perfect detection rates for the StreamSpot dataset (\autoref{table:eval:data:streamspot}).
This makes \unicorn a promising security tool in datacenters and other production environments
that perform well-defined tasks, which are frequently the target of attacks~\cite{equifax, k2010, r2014}.
Hosts that exhibit more diverse behaviors, 
such as workstations~\cite{bernstein2014containers},
pose a greater challenge for IDS in general.
Modern provenance capture systems (\eg CamFlow) help mitigate this issue
as they can separate provenance data based on, \eg namespaces and control groups~\cite{pasquier2017practical}. 
However,
we acknowledge that endpoint security for workstations presents extra challenges that \unicorn does not attempt to address in this work,
as it was originally designed to protect more stable environments.

\noindgras{Larger Cross-evaluation.}
We emphasize that comparing \unicorn with other existing IDS (most of which are syscall-based) is difficult for several reasons:
A) many IDS are not open-source; 
B) existing public IDS datasets are either outdated~\cite{unm, mchugh2000testing} or require a translation~\cite{creech2013generation, haider2016windows, haider2017generating} from, \eg syscall traces to data provenance, 
which is challenging and sometimes impossible (due to lack of information);
C) systems that create their own private datasets only superficially describe their experimental procedures, 
making it difficult to fairly reproduce the experiments for provenance data.
We believe that such a meta-study is a worthwhile endeavor that we plan to pursue in future work. 
 
\section{Related Work}
\label{sec:rw}
This work lies at the intersection of dynamic host-based intrusion detection,
graph-based anomaly detection,
and provenance-based security analysis.
Therefore, we place \unicorn in the context of prior work in these areas.

\noindgras{Dynamic Host-based Intrusion Detection.}
Dynamic host-based intrusion detection (HID) was pioneered by Forrest \etal's
anomaly detection system~\cite{forrest1996sense} that used fixed-length sequences of syscalls to define normal behavior for UNIX processes.
Debar \etal~\cite{debar1998fixed} and Wespi \etal~\cite{wespi2000intrusion} later generalized the approach to incorporate variable-length patterns.
As attacks became increasingly sophisticated~\cite{wagner2002mimicry, tapiador2011masquerade},
systems that modeled only syscall sequences suffered from low detection accuracy.
Next generation systems added state to provide contextual information to the syscalls.
Sekar \etal~\cite{sekar2001fast} designed a finite-state automaton (FSA) that modeled each state as the invocation of a system call from a particular call site.
VtPath~\cite{feng2003anomaly} extended this idea to avoid the impossible path problem, in which there may exist sequences of state transitions in the FSA that cannot happen in practice. 
VtPath performed more extensive call stack analysis to identify such impossible paths as anomalies.
Jafarian \etal~\cite{jafarian2011gray} addressed the impossible path problem by using a deterministic pushdown automata (DPDA).
Maggi \etal~\cite{maggi2010detecting} extended these approaches by combining models of system call sequences with models for the parameters to those system calls.
As automaton-based models approach their theoretical accuracy limit, which is
equivalent to a linear bounded automaton (LBA) or a context-sensitive language model,
their detection capacity increases;
however, the non-polynomial complexity of such theoretical models makes it impossible to realize in practice~\cite{shu2015formal}.
Even a constrained DPDA model exhibits a polynomial time complexity~\cite{jafarian2011gray}.
Shu \etal~\cite{shu2015formal} presented a formal framework that surveyed host-based anomaly detection,
discussing in detail various dynamic and static approaches orthogonal to our work.
Liu \etal~\cite{liu2018host} summarized the state-of-the-art host-based IDS~\cite{khreich2018combining, murtaza2014total, chen2018anrad}
and discussed future research trends, indicating data as one of the decisive factors in IDS research.

\unicorn takes a completely different approach,
because traditional system call approaches are not well-suited for APT attacks (~\autoref{sec:background:syscall})
~\cite{watson2007exploiting, jaeger2004consistency}.
Our graph representation and analysis avoids costly control-flow construction and state transition automata,
while accurately describing and modeling complex relationships among data objects in the system for contextualized anomaly detection.
To the best of our knowledge,
although some systems produce provenance-like graphs from audit logs~\cite{manzoor2016fast}, 
\unicorn is the first system to detect intrusions via 
runtime analysis of native whole-system provenance.

\noindgras{Graph-based Anomaly Detection.}
Akoglu \etal determine graph or subgraph similarity for anomaly detection
by categorizing graphs based on their properties
(\ie \emph{static} vs. \emph{dynamic}, \emph{plain} vs. \emph{attributed})
~\cite{akoglu2015graph}.

Ding \etal~\cite{ding2012intrusion} identified malicious network sources in network flow traffic graphs
based on cut-vertices, using similarity metrics such as betweenness
to detect cross-community communication behavior.
Liu \etal~\cite{liu2005mining} constructed a software behavior graph to describe program execution
and used a support vector machine (SVM) to classify non-crashing bugs (\eg logic errors that do not crash the program)
based on closed subgraphs and frequent subgraphs.
These systems and many other graph mining algorithms~\cite{gao2010community, perozzi2014focused}
and graph similarity measures (\eg graph kernels~\cite{vishwanathan2010graph})
are designed only for static graphs and are difficult to adapt to a streaming setting.

Papadimitriou \etal~\cite{papadimitriou2010web} proposed five similarity schemes for dynamic web graphs, and
NetSimile~\cite{berlingerio2012scalable} used moments of distribution to aggregate egonet-based features (\eg number of neighbors)
to cluster social networks.
Aggarwal \etal~\cite{aggarwal2011outlier} used a structural connectivity model to define outliers and design a reservoir sampling approach
that robustly maintains structural summaries of homogeneous graph streams.
However, these and other streaming-oriented approaches~\cite{mongiovi2013netspot, sun2007graphscope, gupta2011evolutionary}
are either domain specific (\eg bibliographic networks have a different structure than provenance graphs)
or applicable primarily to homogeneous graphs.

In the realm of malware classification and intrusion detection,
Classy~\cite{kostakis2014classy} clusters streams of call graphs to facilitate malware analysis based on graph edit distance (GED)~\cite{gao2010survey} of pairs of graphs
using a modified version of simulated annealing.
Although its runtime complexity is suitable for graph streams,
the empirical evaluation was limited to graphs with no more than 3,000 vertices;
real system execution yields graphs orders of magnitude larger~\cite{pasquier2018ccs}.

StreamSpot~\cite{manzoor2016fast} analyzes streaming information flow graphs to detect anomalous activity.
However, StreamSpot's graph features are locally constrained while \unicorn's embody execution context.
We show in~\autoref{sec:evaluation} that contextualized graph analysis has great impact on detection performance.
Furthermore, StreamSpot models only a single snapshot of every training graph,
dynamically maintaining its clusters during test time.
However,
it results in a significant number of false alarms, which creates an opportune time window for the attacker.
We also consider such an approach inappropriate in APT scenarios,
where persistent attackers can manipulate the model to gradually and slowly change system behavior to avoid detection.
\unicorn takes full advantage of its ability to continuously summarize the evolving graph,
modeling the corresponding evolution of the system execution it monitors.
FRAPpuccino~\cite{han2017frappuccino} is another attempt at graph-based intrusion detection.
It uses a windowing approach to allow for efficient graph analysis.
Naturally, segmenting provenance graphs in this way produces a more limited view of system execution, 
unsuitable for long-term detection that spans windows.

\noindgras{Provenance-based Security Analysis}
A variety of security-related applications leverage provenance, 
mostly notably for forensic analysis and attack attribution~\cite{akoglu2015graph}.
BackTracker~\cite{king2003backtracking} analyzes intrusions using a provenance graph to identify the entry point of the intrusion,
while PriorTracker~\cite{liu2018towards} optimizes this process and enables a forward tracking capability for timely attack causality analysis.
HERCULE~\cite{pei2016hercule} analyzes intrusions by discovering attack communities embedded within provenance graphs.
Winnower~\cite{hassan2018towards} expedites system intrusion investigation through grammatical inference over provenance graphs,
and simultaneously reduces storage and network overhead without compromising the quality of provenance data.
NoDoze~\cite{hassannodoze} performs attack triage within provenance graphs to identify anomalous paths.
Bates \etal~\cite{bates2015trustworthy} were the first to use provenance for data loss prevention,
and Park \etal~\cite{park2012provenance} formalized the notion of provenance-based access control (PBAC).
Ma \etal~\cite{ma2016protracer} designed a lightweight provenance tracing system ProTracer 
to mitigate the dependence explosion problem and reduce space and runtime overhead, 
facilitating practical provenance-based attack investigation.
Pasquier \etal~\cite{pasquier2018ccs} introduced a generic framework, called CamQuery, that enables inline, realtime provenance analysis,
demonstrating great potential for future provenance-based security applications.

Recently,
as APT attacks become increasingly prominent, 
a number of systems leverage data provenance for APT attack analysis.
Holmes~\cite{milajerdi2019holmes} and 
Sleuth~\cite{hossain2017sleuth} focus primarily on attack reconstruction using information flow provided by data provenance.
Their approach is similar to an architecture
proposed by Tariq \etal~\cite{tariq2011identifying} that correlates anomalous activity in grid applications using data provenance.
The anomaly detection module itself uses a simple, pre-defined model
that relies on expert knowledge of the existing APT kill-chain
to match an \emph{a priori} specification of possible exploits to localized components in the graph.
Poirot~\cite{milajerdi2019poirot} produces another form of attack reconstruction. 
It correlates a collection of compromise indicators (found by other systems) to identify APTs.
Relying on expert knowledge from existing cyber threat intelligence reports and compromise descriptions
to construct attack query graphs,
Poirot performs offline graph pattern matching on provenance graphs to uncover potential APTs.
For example, it uses the red team's attack descriptions to manually craft 
query graphs to correlate anomalies in the DARPA datasets used in \autoref{sec:evaluation:darpa}.
This is a critical limitation given that 
composing a sufficiently detailed description of a new class of APT takes significant forensic effort~\cite{barre2019mining}.

\unicorn differs from these rule-based systems in that it is an anomaly-based system that 
requires no prior expert knowledge of APT attack patterns and behaviors.
Although rule-based approaches are closely aligned with commercial practices today
(\ie they are essentially provenance-based versions of the Endpoint Detection and Response (EDR) tools 
offered by enterprise security vendors),
prior research shows that rule-based EDR systems are the chief contributor to the ``threat fatigue problem''~\cite{fireeye}.
Recent work on provenance analysis (\eg NoDoze~\cite{hassannodoze}) demonstrated 
that historical context is crucial for mitigating this problem; 
\unicorn instead investigates how to incorporate context as a first class citizen in HID systems, 
rather than as a secondary triage tool.

Gao \etal~\cite{gao2018saql} leveraged complex event processing platforms and designed a domain-specific query language, SAQL to 
analyze large-scale, streaming provenance data.
The system combines various anomaly models (\eg rule-based anomalies)
and aggregates data streams across multiple hosts,
but it ultimately requires expert domain knowledge to identify elements/patterns to match against queries.
We also note that our provenance graph analyses are able to incorporate implicitly (without domain knowledge) most of their anomaly models (\eg invariant-based, time-series, and outlier-based).
Barre \etal~\cite{barre2019mining}
mine data provenance to detect anomalies.
The goal of their work is mainly to identify important process features that are likely to be relevant to APT attacks (\eg a process' lifetime and path information).
Using a random forest model with hand-picked process features,
their system delivers a detection rate of only around 50\%.
Such low performance suggests that simple feature engineering on provenance graphs
without considering graph topology is insufficient in detecting stealthy APT attacks.
Berrada \etal~\cite{berrada2019aggregating} proposes score aggregation techniques
to combine anomaly scores from different anomaly detectors to improve detection performance.
Although their work targets provenance graphs for APT detection,
it is orthogonal to \unicorn (or any other detectors) in that it functions only as an aggregator for existing anomaly detection systems.
 
\section{Conclusion}
\label{sec:conclusion}
We present \unicorn, a realtime anomaly detection system that leverages whole-system data provenance
to detect advanced persistent threats that are deemed difficult for traditional detection systems.
\unicorn models system behavior via structured provenance graphs that expose causality relationships between system objects,
taking into account the entirety of the graph by efficiently summarizing it as it streams into its analytic pipeline.
Our evaluation shows that the resulting evolutionary models can successfully detect various APT attacks captured from different audit systems,
including real-life APT campaigns, with high accuracy and low false alarm rates.
 
\section*{Acknowledgement}
\noindent We thank Dr. Robert N. W. Watson and his team at the University of Cambridge on the DARPA Transparent Computing Program.
We also thank the anonymous reviewers and our shepherd Prof. Brendan Saltaformaggio who helped improve the paper.
This research was supported by the US National Science Foundation under grants NSF 14-50277, 16-57534, and 17-50024.
This research was also supported with promotional credits from the AWS Cloud Credits for Research program.
The views, opinions, and/or findings contained in this paper are those of the authors and should not be interpreted as representing the official views or policies, either expressed or implied, of the Department of Defense or the U.S. Government.
 
\footnotesize{\bibliographystyle{IEEEtranS}}
\bibliography{biblio}

\appendix

\subsection{Availability}
\label{sec:availability}
\noindent The implementation described in \autoref{sec:implementation} and the material to reproduce the evaluation presented in \autoref{sec:evaluation} are available online under Apache License V2.0 and GPL v2 (see individual subcomponents for more details) at \url{https://github.com/crimson-unicorn}.
 
\subsection{HistoSketch}
\label{sec:app:histosketch}
\noindgras{Notation and Basic Concepts.}
\autoref{tab:notations} presents the notation we use in the rest of this section.
Similarity-preserving data sketching, or locality sensitive hashing (LSH),
allows us to efficiently compute the similarity between two graphs by projecting their high-dimensional histogram representations
to a low-dimensional space while preserving their similarity~\cite{manzoor2016fast}.

\begin{table}[h!]\begin{center}
\begin{tabular}{l c l}
\toprule
$\epsilon$ & $\triangleq$ & Set of histogram elements\\
$h$ & $\triangleq$ & Histogram element: $h \in \epsilon$\\
$L$ & $\triangleq$ & Cumulative, weighted histogram count vector: $L \in \mathbb{R}_{>0}^{|\epsilon|}$\\
$L_h$ & $\triangleq$ & Count of histogram element $h$\\
$\lambda$ & $\triangleq$ & Weight decay factor\\
$w_t$ & $\triangleq$ & Exponential decay weight: $w_t = e^{-\lambda\Delta t}$\\
$S$ & $\triangleq$ & Sketch of histogram vector $L$: $|S| \ll |\epsilon|$ and is fixed\\
\bottomrule
\end{tabular}
\end{center}
\caption{Notation}
\label{tab:notations}
\end{table}

\noindent Formally, we define LSH as follows (adopted from Charikar~\cite{charikar2002similarity}):
\begin{definitionii}
A locality sensitive hashing scheme is a distribution on a family $\mathcal{F}$ of hash functions operating on a collection of objects,
such that for two objects $m$, $n$,
{\scriptsize
\begin{align}
P_{h\in\mathcal{F}}[h(m) = h(n)] = sim(m, n) \nonumber
\end{align}
}
where $sim(m, n)$ is some similarity function defined on the collection of objects.
\end{definitionii}

HistoSketch~\cite{yang2017histosketch} uses \emph{normalized min-max similarity} to measure the similarity between two histogram vectors:
\begin{definitionii}
{\scriptsize
\begin{align}
sim_{min-max}(H^{a}, H^{b}) &= \frac{\sum_{h\in\epsilon}min(H_{h}^{a}, H_{h}^{b})}{\sum_{h\in\epsilon}max(H_{h}^{a}, H_{h}^{b})} \nonumber \\
\sum_{h\in\epsilon}H_{h}^{a} &= 1 \nonumber \\
\sum_{h\in\epsilon}H_{h}^{b} &= 1 \nonumber
\end{align}
}
where the superscript $a$, $b$ denotes the identity of a histogram.
\end{definitionii}

\noindgras{Sketch Creation.}
HistoSketch uses a variation of consistent weighted sampling that takes as input positive real numbers
to generate fixed-size sketches~\cite{li20150}.
The size of the sketch $|S|$ controls the tradeoffs between information loss and computation efficiency for real-time detection (\autoref{sec:evaluation}).

To generate one sketch element $S_j$,
we first draw three random variables for each $h \in \epsilon$:
{\scriptsize
\begin{align}
r_{h, j} &\sim Gamma(2, 1) \nonumber \\
c_{h, j} &\sim Gamma(2, 1) \nonumber \\
\beta_{h, j} &\sim Uniform(0, 1) \nonumber
\end{align}
}
and then follow the steps in \autoref{algo:sketch}.
\begin{algorithm}
	\scriptsize
	\algsetup{linenosize=\tiny}
	\SetAlgoLined
	\DontPrintSemicolon
	\SetKwInOut{Input}{Input}\SetKwInOut{Output}{Output}
	\Input{Histogram $L$, $r, c, \beta$}
	\Output{Sketch $S$ and the corresponding hash values $A$}
	\For{$j \leftarrow 1$ \KwTo $|S|$}{
		\ForEach{$h \in \epsilon$}{
			$y_{h, j} = exp(logL_{h} - r_{h, j}\beta_{h, j})$\;
			$a_{h, j} = \frac{c_{h, j}}{y_{h, j}exp(r_{h, j})}$\;
		}
		$S_{j} = argmin_{h\in\epsilon}a_{h, j}$\;
		$A_{j} = min_{h\in\epsilon}a_{h, j}$\;
	}
	\caption{Creating Graph Sketch Using HistoSketch}\label{algo:sketch}
\end{algorithm}
$r$, $c$, and $\beta$ are fixed for each element in $\epsilon$.
The sketch element $j$ is the element $h$ in $\epsilon$ whose hashed value $a_{h, j}$ is the minimum in the $j^{th}$ column of matrix $A$.

\noindgras{Sketch Update.}
HistoSketch makes it possible to quickly update the sketch as new data arrives.
At time $t + 1$, it incrementally updates the sketch $S(t+1)$ based on the weighted histogram $L$,
the previous sketch $S(t)$ and its corresponding hash values $A(t)$,
the new data item $x_{t+1}$, and the weight decay factor $\lambda$.
\autoref{algo:update} describes this process in detail.

\begin{algorithm}
	\scriptsize
	\algsetup{linenosize=\tiny}
	\SetAlgoLined
	\DontPrintSemicolon
	\SetKwInOut{Input}{Input}\SetKwInOut{Output}{Output}
	\Input{$L$, $S(t)$, $A(t)$, $x_{t+1}$, $\lambda$}
	\Output{New sketch $S(t+1)$ and the corresponding hash values $A(t+1)$}
	\ForEach{$h \in \epsilon$}{
		$L_{h} = L_{h}(t)\cdot e^{-\lambda}$\;\tcc*[r]{exponential decay}
	}
	\For{$j \leftarrow 1$ \KwTo $|S|$}{
		\lIf{$x_{t+1} \in \epsilon$}{$L_{x_{t+1}} = L_{x_{t+1}} + 1$}
		\Else{
			$\epsilon = \epsilon + \{x_{t+1}\}$\;
			$L_{x_{t+1}} = 1$\;
		}
		Compute $a_{x_{t+1}, j}$\; \tcc*[r]{see \autoref{algo:sketch}}
		\If{$a_{x_{t+1}, j} < A_{j}(t)\cdot e^{-\lambda}$}{
			$S_{j}(t+1) = x_{t+1}$\;
			$A_{j}(t+1) = a_{x_{t+1}, j}$\;
		}
		\Else{
			$S_{j}(t+1) = S_{j}(t)$\;
			$A_{j}(t+1) = A_{j}(t)\cdot e^{-\lambda}$\;
		}
	}
	\caption{Updating Graph Sketch}\label{algo:update}
\end{algorithm}
 
\subsection{Metrics}
\label{sec:metrics}
We denote false positives by $fp$,
false negatives $fn$, true positives $tp$, and true negatives $tn$.

\begin{definitionii}
precision = $\frac{tp}{tp+fp}$	
\end{definitionii}
\begin{definitionii}
	recall = $\frac{tp}{tp+fn}$
\end{definitionii}
Precision and recall measure \emph{relevance} in classification,
where precision is a measure of \emph{exactness} and recall \emph{completeness}.

\begin{definitionii}
	accuracy = $\frac{tp+tn}{tp+tn+fp+fn}$
\end{definitionii}
\begin{definitionii}
	F-score = $2\times\frac{precision\times recall}{precision + recall}$
\end{definitionii}
F-score combines precision and recall using their harmonic mean.
In our measurement, recall and precision are evenly weighted. \end{document}